\documentclass[twocolumn,english,showpacs,preprintnumbers,superscriptaddress,nofootinbib,prb]{revtex4-1}
\usepackage[latin9]{inputenc}
\usepackage[english]{babel}
\usepackage{setspace}
\usepackage{graphicx}
\usepackage{amssymb}
\usepackage{amsmath}
\usepackage{amscd}
\usepackage{rotating}
\usepackage[usenames, dvipsnames]{color}
\usepackage{epstopdf}
\usepackage{comment}
\usepackage{siunitx}
\usepackage{float}
\usepackage{xfrac}
\usepackage{natbib}
\usepackage{hyperref}
\hypersetup{
    colorlinks=true,
    linkcolor=blue,
    filecolor=blue,
    urlcolor=blue,
   citecolor=blue,
}
\usepackage[draft]{todonotes}

\usepackage[normalem]{ulem}
\newcommand\footnoteref[1]{\protected@xdef\@thefnmark{\ref{#1}}\@footnotemark}

\begin{document}

\title{Spin-dependent zero-bias peak in a hybrid nanowire-quantum dot system: Distinguishing isolated Majorana fermions from Andreev bound states }

\author{L. S. Ricco}
\email[corresponding author: ]{luciano.silianoricco@gmail.com}
\affiliation{Departamento de F\'{i}sica e Qu\'{i}mica, Universidade Estadual Paulista (Unesp), Faculdade de Engenharia, 15385-000, Ilha Solteira, S\~ao Paulo, Brazil}
\author{M. de Souza}
\affiliation{Departamento de F\'{i}sica, Instituto de Geoci\^encias e Ci\^encias Exatas, Universidade Estadual Paulista (Unesp), 13506-970, Rio Claro, S\~ao Paulo, Brazil}
\author{M. S. Figueira}
\affiliation{Instituto de F\'isica, Universidade Federal Fluminense, 24210-340, Niter\'oi, Rio de Janeiro, Brazil}
\author{I. A. Shelykh}
\affiliation{Science Institute, University of Iceland, Dunhagi-3, IS-107,
Reykjavik, Iceland}
\affiliation{ITMO University, St. Petersburg 197101, Russia}
\author{A. C. Seridonio}
\email[corresponding author: ]{acfseridonio@gmail.com}
\affiliation{Departamento de F\'{i}sica e Qu\'{i}mica, Universidade Estadual Paulista (Unesp), Faculdade de Engenharia, 15385-000, Ilha Solteira, S\~ao Paulo, Brazil}
\affiliation{Departamento de F\'{i}sica, Instituto de Geoci\^encias e Ci\^encias Exatas, Universidade Estadual Paulista (Unesp), 13506-970, Rio Claro, S\~ao Paulo, Brazil}

\date{\today}

\begin{abstract}
Hybrid system composed by a semiconducting nanowire with proximity-induced superconductivity and a quantum dot at the end working as spectrometer was recently used to quantify the so-called degree of Majorana nonlocality [Deng \textit{et al.,} \href{https://journals.aps.org/prb/abstract/10.1103/PhysRevB.98.085125}{Phys.Rev.B, \textbf{98}, 085125 (2018)}]. Here we demonstrate that spin-resolved density of states of the dot responsible for zero-bias conductance peak strongly depends on the separation between the Majorana bound states (MBSs) and their relative couplings with the dot and investigate how the charging energy affects the spectrum of the system in the distinct scenarios of Majorana nonlocality (topological quality). Our findings suggest that spin-resolved spectroscopy of the local density of states of the dot can be used as a powerful tool for discriminating between different scenarios of the emergence of zero-bias conductance peak.

\end{abstract}
\maketitle

\section{Introduction}

The possibility of achieving of fault-tolerant quantum computing with qubits based on Majorana bound states (MBSs)~\cite{ReviewAguado,kitaev} started a new era in the domains of mesoscopic physics and quantum information. These exotic non-Abelian excitations~\cite{RevNonabelian} emerge as topologically protected mid-gap zero-energy modes in so-called topological superconductors~\cite{Alicea,RevMajorana}. The topological protection stems from the separation between individual MBSs, i.e, \textit{nonlocality}, which is also responsible for the immunity of a setup against local perturbations and consequent loss of the information due to the processes of decoherence~\cite{ReviewDasSarma,RevNonabelian}. However, it should be noticed that, for practical realizations of quantum computing systems, the MBSs qubit becomes vulnerable to the decoherence process caused by local perturbations when coupled to environment~\cite{Decaymemories,Failureprotection}, which can lead to unwelcome errors in the processing of quantum information.    

Topological superconductivity can be realized experimentally in hybrid superconductor-semiconductor nanowires with induced proximity effect in the presence of strong spin-orbit coupling and external magnetic field, favoring the formation of superconducting (SC) triplet states ~\cite{Nanowire1,Nanowire2}. In these hybrid devices, manifestation of a robust zero-bias conductance peak (ZBCP) has been considered as an experimental signature of the presence of highly nonlocal MBSs emerging at the opposite ends of a nanowire \cite{mourik2012,albrecht2016exponential,Deng2016,AguadoExperiment,Quantized}. However, it was argued later on that other physical mechanisms such as disorder~\cite{Disorder}, Kondo effect~\cite{Kondo1,Kondo2} and formation of Andreev bound states (ABSs)~\cite{ABS1,ABSLee,ABS2,psABS1,psABS2,Cayao2018,YSR,SNSjunction,Josepheson,SanJose2013} can be responsible for the appearance of analogs of ZBCP. In particular, there is ongoing controversy~\cite{nonHermitan} whether near zero-energy ABS, constituted by weakly overlapping MBSs, can mimic robust $2e^{2}/h$ ZBCP ~\cite{psABS1,psABS2}. Overall, there is currently consensus that observation of ZBCP only is not enough to guarantee the presence of topologically protected MBSs in the system.
\begin{figure}[t]
	\centerline{\includegraphics[width=3.5in,keepaspectratio]{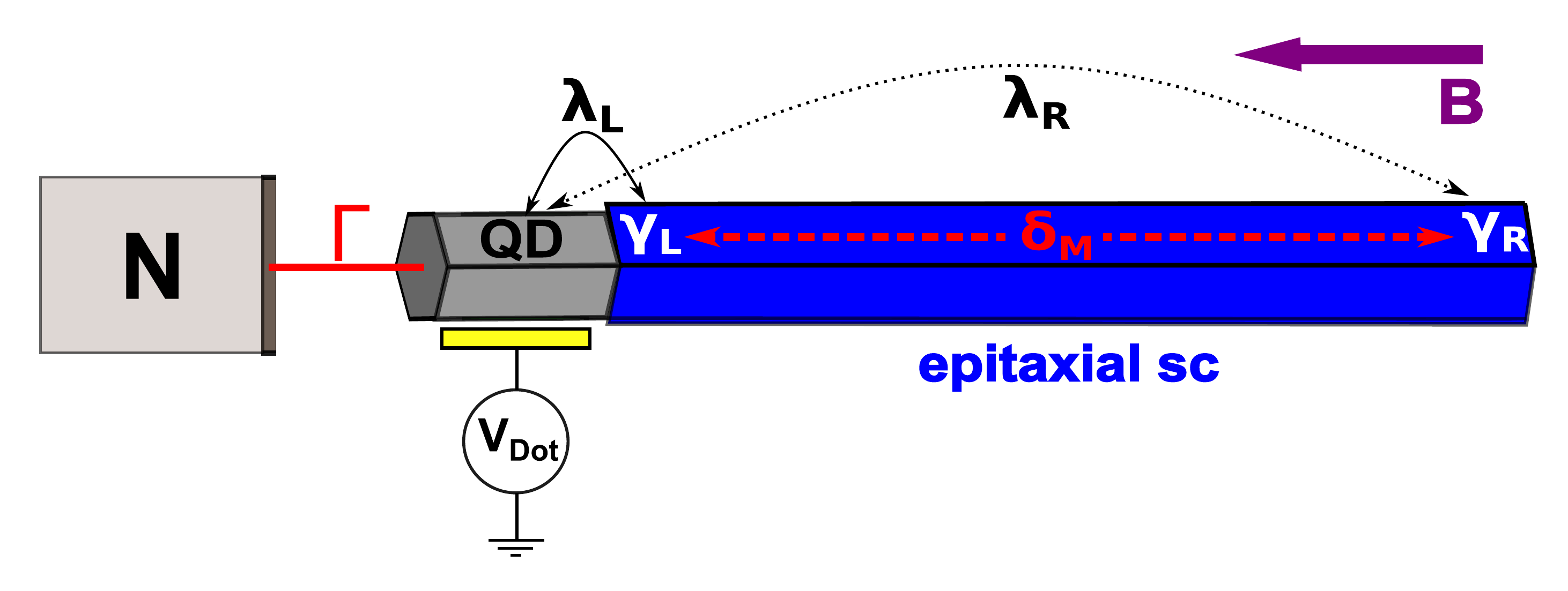}}
	\caption{Sketch of the system consisting of a hybrid superconducting (SC) nanowire (blue region) coupled to a quantum dot (QD) with energy level $\varepsilon_{d}$, which can be tuned by application of an external gate voltage $V_{\text{Dot}}$. The QD is coupled to both Majorana bound states (MBSs) $\gamma_{L}$ and $\gamma_{R}$ at the opposite ends of SC nanowire with strengths $\lambda_{L}$ and $\lambda_{R}$, respectively. The MBSs may be hybridized with each other by $\delta_{M}$ in the  presence of an external magnetic field applied longitudinally (purple arrow) due to the finite size effects. The QD levels are broadened due to the coupling $\Gamma$ with a normal metallic lead N.
		\label{Setup}}
\end{figure}

A possible way to clarify the origin of ZBCP is performing tunneling spectroscopy of a quantum dot (QD)~\cite{Hoffman,Denis} is assumed to be coupled to both ends of SC nanowire. In this type of experiment one can access the so-called degree of Majorana nonlocality~\cite{RAguado,DJClarcke,AguadoExperiment} characterizing \textit{``how topological''} are MBSs and distinguish between the cases of well-separated MBS and near zero-energy ABSs (overlapping MBSs)~\cite{Deng2016,ABS2,ABS3,ABS4}. 

Distinct from earlier works ~\cite{Deng2016,RAguado,DJClarcke,AguadoExperiment,MBSoscillations}, in the present paper we analyze how charging energy of the QD single-level coupled to a normal lead affects the energy spectrum of the device sketched in the Fig.~\ref{Setup}. To account for the correlation effects we go beyond the Hartree-Fock mean-field approximation used by Prada \textit{et al}~\cite{RAguado} for larger Zeeman fields, applying the method developed by Hubbard~\cite{HubbardI} to treat the charging energy which allows us to recover qualitatively the recent experimental profiles reported by Deng \textit{et al}~\cite{AguadoExperiment} for a set of tunable parameters. In this sense, we demonstrate that for highly nonlocal MBSs, a plateau at zero-energy is formed in the QD density of states in the wide range of the values of the dot level and charging energies. In the case of strongly overlapping MBSs forming an ABS, the spectrum is strongly modified and this plateau disappears. Moreover, changes in MBSs degree of nonlocality strongly affect spin resolved density of states of QD, which means that spectroscopic experiment with spin-polarized local probe~\cite{Bena,Bjornson,Tadeusz,Jeon,Andrei,Jelenaspinpolarized} will allow to identify whether ZBCP is induced by topological MBSs or ABSs (overlapping Majoranas).
 \section{The Model}
 
 The effective Hamiltonian describing the device depicted in Fig.~\ref{Setup} takes the following form \cite{RAguado,MBSoscillations}
\begin{eqnarray}
H&=&\sum_{\mathbf{k}\sigma}\varepsilon_{\mathbf{k}\sigma}e_{\mathbf{k}\sigma}^{\dagger}e_{\mathbf{k}\sigma}+\sum_{\sigma}\varepsilon_{d\sigma}n_{d\sigma}+Un_{d\uparrow}n_{d\downarrow} \nonumber \\
& + & \sqrt{2}V\sum_{\mathbf{k}\sigma}(e_{\mathbf{k}\sigma}^{\dagger}d_{\sigma}+d_{\sigma}^{\dagger}e_{\mathbf{k}\sigma}) + H_{Nw},\label{eq:FullH}
\end{eqnarray}
where $n_{d\sigma}=d_{\sigma}^{\dagger}d_{\sigma}$ is the operator of the number of the electrons residing in the single-level QD, the operators  $e_{\mathbf{k}\sigma}^{\dagger}(e_{\mathbf{k}\sigma})$ correspond to the creation of the electrons (holes) in the normal (N) lead~\cite{MBSoscillations} with wave vector $\mathbf{k}$, spin $\sigma=\uparrow,\downarrow$ and energy $\varepsilon_{\mathbf{k}\sigma}$. The energy of an electron in the QD is spin dependent, $\varepsilon_{d\sigma}=\varepsilon_{d}-\sigma V_{Z}$, where
$V_{Z}$ is Zeeman energy splitting induced by an external magnetic field. $U$ is the charging energy of the QD. The dot is coupled to  the normal lead N with coupling strength $\sqrt{2}V$.

To describe the SC-nanowire hosting a pair of MBSs $\gamma_{i}$ at the opposite ends and coupled to the QD, we use low-energy effective model developed by Prada \textit{et. al}~\cite{RAguado}, and characterized by the following Hamiltonian:
\begin{eqnarray}
H_{Nw}&=&\imath\delta_{M}\gamma_{L}\gamma_{R}+(\lambda_{L}d_{\sigma}-\lambda_{L}^{*}d_{\sigma}^{\dagger})\gamma_{L}\nonumber \\ &+& (\lambda_{R}d_{\sigma} + \lambda_{R}^{*}d_{\sigma}^{\dagger})\gamma_{R}\label{eq:HNw}
\end{eqnarray}
where self-conjugated  operators $\gamma_{i}=\gamma_{i}^{\dagger}$ describe localized Majorana fermions and obey the algebraic relation $\{ \gamma_{i}, \gamma_{j} \}  = \delta_{ij}$~\cite{ReviewAguado,Alicea,ReviewDasSarma,RevMajorana}. $H_{Nw}$ can be rewritten in the regular fermionic basis, since Majorana operators can be decomposed into pairs of normal fermionic operators, $\gamma_{L}=(f_{\uparrow} + f_{\uparrow}^{\dagger})/\sqrt{2}$  and $\gamma_{R}=\imath(f_{\uparrow}^{\dagger}-f_{\uparrow})/\sqrt{2}$. The dot is coupled to the left and right MBSs, with coupling constants $\lambda_{L}$ and $\lambda_{R}$, respectively. The direct  hybridization $\delta_{M}$ between MBSs reads~\cite{Danon,Fleckenstein}
\begin{equation}
\delta_{M}=\frac{e^{-{l/2b}}}{\sqrt{b}}\cos(l\sqrt{b})E_{0}, \label{eq:deltaM}
\end{equation}
which is the function of both Zeeman energy splitting $b=V_{Z}/E_{0}$, $E_{0}=\left(2m^{*}\alpha^{2}\Delta_{\text{SC}}^{2}/\hbar^{2}\right)^{1/3}$ and $l=L\sqrt{2m^{*}E_{0}}/\hbar$ with $L$ being the length of the wire, $m^{*}$  being electrons effective mass, $\alpha$ the spin-orbit coupling constant and $\Delta_{\text{SC}}$ the induced SC gap~\cite{Danon}. 
The degree of MBSs nonlocality $\eta$ can be defined as ratio between QD-MBSs right/left coupling strengths~\cite{RAguado}:
\begin{equation}
\eta^{2} = \frac{|\lambda_{R}|}{|\lambda_{L}|}
\end{equation}
This parameter can be experimentally accessed through the measurement of the conductance as a function of the gate potential changing the energy of a QD and  drain-source voltage~\cite{AguadoExperiment} and estimated as the ratio between energy values in which the Majorana and QD states are on resonance (anticrossing points)~\cite{RAguado}, $\eta^2 \approx \epsilon_{\text{MBS}}^{\pm}/\epsilon_{\text{QD}}^{\pm}$ [See Fig~\ref{Fig2}(d-f)]. 
\subsection{Density of states calculations} 

Our main goal is to investigate how the spectral properties of the QD accessible in spin-resolved measurements are changing when the degree of MBSs nonlocality characterized by the parameter $\eta$~\cite{RAguado} is modified. Hence, it is suitable to evaluate the total density of states (DOS) in the QD, which reads:
\begin{equation}
\text{DOS}(\omega) = \pi\Gamma \sum_{\sigma}\rho_{\sigma}(\omega),\label{eq:DOS}
\end{equation}
where the constant $\Gamma = 2\pi V^{2} \rho_{0}$ is the QD-N lead effective coupling~\cite{Anderson}, with $\rho_{0}$ being the DOS of the lead. The quantity
\begin{equation}
 \rho_{\sigma}(\omega)=-\frac{1}{\pi}\text{Im}[G^{r,\sigma}_{d,d}(\omega)]   
\end{equation}
denotes the DOS corresponding to a given spin orientation, which is determined by the retarded Green's function of the QD $G^{r,\sigma}_{d,d}(\omega)$ in the spectral domain. The application of the equation of motion (EOM) method~\cite{Jauho} leads to the following equation \textcolor{blue}{(Appendix \ref{AppendixA}):}
\begin{eqnarray}
(\omega^{+}-\varepsilon_{d\sigma}-\Sigma_{\text{M},\sigma}^{\text{U=0}}+\imath\Gamma)G_{d,d}^{r,\sigma}(\omega)=1+UG_{d_{\sigma}n_{d\bar{\sigma}},d_{\sigma}}^{r}(\omega)\nonumber\\+U(|\lambda_{L}|^{2}-|\lambda_{R}|^{2})\bar{K}^{\sigma}G_{d_{\sigma}^{\dagger}n_{d\bar{\sigma}},d_{\sigma}}^{r}(\omega),\nonumber\\
\label{eq:E1}
\end{eqnarray}
where $\omega^+=\omega+i0^{+}$, $\Sigma_{\text{M},\sigma}^{\text{U=0}}=K_{1} +(|\lambda_{L}|^{2}-|\lambda_{R}|^{2})^{2}K\bar{K}^{\sigma}$ is the self-energy~\cite{MBSoscillations} due to QD-MBSs hybridization in the absence of the charging energy, and 
\begin{equation}
K=\frac{1}{2}\left(\frac{1}{\omega^{+}+\delta_{M}}+\frac{1}{\omega^{+}-\delta_{M}}\right)\label{eq:K},    
\end{equation}
%
\begin{equation}
\bar{K}^{\sigma} = \frac{1}{2} \left(\frac{K}{\omega^{+} + \varepsilon_{d\sigma}-K_{2} + \imath \Gamma}\right),\label{eq:Kbar}
\end{equation}
\begin{equation}
K_{1} = \frac{1}{2}\cdot \left[\frac{(|\lambda_{L}|-|\lambda_{R}|)^{2}}{\omega^{+}-\delta_{M}} + \frac{(|\lambda_{L}|+|\lambda_{R}|)^{2}}{\omega^{+}+\delta_{M}}\right] \label{eq:K1}
\end{equation}
and
\begin{equation}
K_{2} = \frac{1}{2}\cdot\left[\frac{(|\lambda_{L}|-|\lambda_{R}|)^{2}}{\omega^{+}+\delta_{M}} + \frac{(|\lambda_{L}|+|\lambda_{R}|)^{2}}{\omega^{+}-\delta_{M}}\right]. \label{eq:K2}
\end{equation}
The presence of the two-particle operator corresponding to the charging energy term in the  Hamiltonian [Eq.~(\ref{eq:FullH})] leads to the appearance of the two particle Green's functions in the Eq.~(\ref{eq:E1}). The iterative application  of the EOM procedure to such higher order functions will produce an infinite chain of the equations which should be truncated at some point~\cite{Jauho}. Distinct from the earlier work~\cite{RAguado} in which the charging energy of the dot was accounted for using a mean-field approximation, here we take a step further by following Hubbard-I truncation scheme~\cite{HubbardI}.  This allows us to account for the appearance of the so-called Hubbard peaks and thus describe better the physics of Coulomb blockade regime. Note, however, that in our approach Kondo-type correlations are fully neglected, and it is applicable only if $T_{K}/\Delta_{\text{SC}}\gtrapprox0.6$~\cite{RAguado} or $T\gg T_{K}$, wherein $T_{K}$ is the Kondo temperature~\cite{Kondo1,Kondo2}. Further details of the calculations can be found in the Appendix \ref{AppendixB}.  

After Hubbard-I truncation, the two-particle Green's functions take the following form:
\begin{equation}
G_{d_{\sigma}n_{d\bar{\sigma}};d_{\sigma}}^{r}(\omega)=\frac{\langle n_{d\bar{\sigma}}\rangle}{\omega^{+}-\varepsilon_{d\sigma}-U-\Sigma_{\text{M},\sigma}^{\text{U\ensuremath{\neq}0}}+\imath \Gamma},\label{eq:E2}
\end{equation}
and
\begin{equation}
G_{d_{\sigma}^{\dagger}n_{d\bar{\sigma}}d_{\sigma}}^{r}(\omega)=-(|\lambda_{L}|^{2}-|\lambda_{R}|^{2})\bar{K}_{\text{U}}^{\sigma}G_{d_{\sigma}n_{d\bar{\sigma}},d_{\sigma}}^{r}(\omega),\label{eq:E3}
\end{equation}
wherein
\begin{equation}
\langle n_{d\bar{\sigma}}\rangle = \int_{-\infty}^{0} d\omega\rho_{\bar{\sigma}}(\omega) \label{eq:occupation}
\end{equation}
gives the occupation number of the dot per spin $\bar{\sigma}$ (opposite to $\sigma$) at $T=0$. The self-energy term provided by the presence of MBSs and charging energy $U$ is given by
\begin{equation}
\Sigma_{\text{M},\sigma}^{\text{U\ensuremath{\neq}0}}=K_{1}+(|\lambda_{L}|^{2}-|\lambda_{R}|^{2})^{2}K\bar{K}_{\text{U}}^{\sigma}    
\end{equation}
where
\begin{equation}
\bar{K}_{\text{U}}^{\sigma}=\frac{1}{2}\cdot\frac{K}{\omega^{+}+\varepsilon_{d\sigma}+U+\-K_{2} + \imath \Gamma}
\end{equation}

After some algebra we get from Eqs.~(\ref{eq:E1}), (\ref{eq:E2}) and (\ref{eq:E3}) the following expression for the retarded Green's function of the dot:
\begin{equation}
G_{d,d}^{r,\sigma}(\omega)=\frac{\lambda(\omega,\sigma\bar{\sigma})-U(|\lambda_{L}|^{2}-|\lambda_{R}|^{2})^{2}\mathcal{M}(\omega,\sigma\bar{\sigma})}{\omega^{+}-\varepsilon_{d\sigma}-\Sigma_{\text{M},\sigma}^{\text{U=0}} + \imath \Gamma}\label{eq:Gddfinal}
\end{equation}
with
\begin{equation}
\lambda(\omega,\sigma\bar{\sigma})=1+\frac{U\langle n_{d\bar{\sigma}}\rangle}{\omega^{+}-\varepsilon_{d\sigma}-U-\Sigma_{\text{M},\sigma}^{\text{U\ensuremath{\neq}0}} + \imath \Gamma}\label{eq:lambda}
\end{equation}
and
\begin{equation}
\mathcal{M}(\omega,\sigma\bar{\sigma})= \frac{\langle n_{d\bar{\sigma}}\rangle\bar{K}^{\sigma}\bar{K}_{\text{U}}^{\sigma}}{\omega^{+}-\varepsilon_{d\sigma}-U-\Sigma_{\text{M},\sigma}^{\text{U\ensuremath{\neq}0}}+\imath \Gamma}. \label{eq:M}
\end{equation}
 \section{Results and Discussion}
 
 We investigate the energy spectrum of the device depicted in Fig.~\ref{Setup} analyzing the DOS of the QD [Eq.~(\ref{eq:DOS})] as a function of spectral frequency $\omega$ and dot energy level $\varepsilon_{d}$ for several regimes corresponding to the different ratios between the parameters of the system. The relevant parameters of our model, in units of $E_{0}$, are the charging energy $U$, hybridization $\lambda_{L}(\lambda_{R})$ between the dot and MBS(left/right) and Zeeman energy splitting $V_{Z}$, which modulates the direct overlap $\delta_{M}$ between the MBSs at the opposite nanowire ends. The length of the SC was chosen as $L\approx0.1l\si{\micro\metre}$, in accordance with the results presented in the Ref.~\onlinecite{Danon}. The occupation numbers for each spin [Eq.~(\ref{eq:occupation})] were self-consistently computed. In all the situations, the QD-left MBS coupling strength is kept fixed ($\lambda_{L}=1.0E_{0}$). Concerning the charging energy strength $U$, we follow Prada \textit{et al.}~\cite{RAguado} effective Hamiltonian, assuming that $U>\Delta$.

\begin{figure}[t]
	\centerline{\includegraphics[width=3.6in,keepaspectratio]{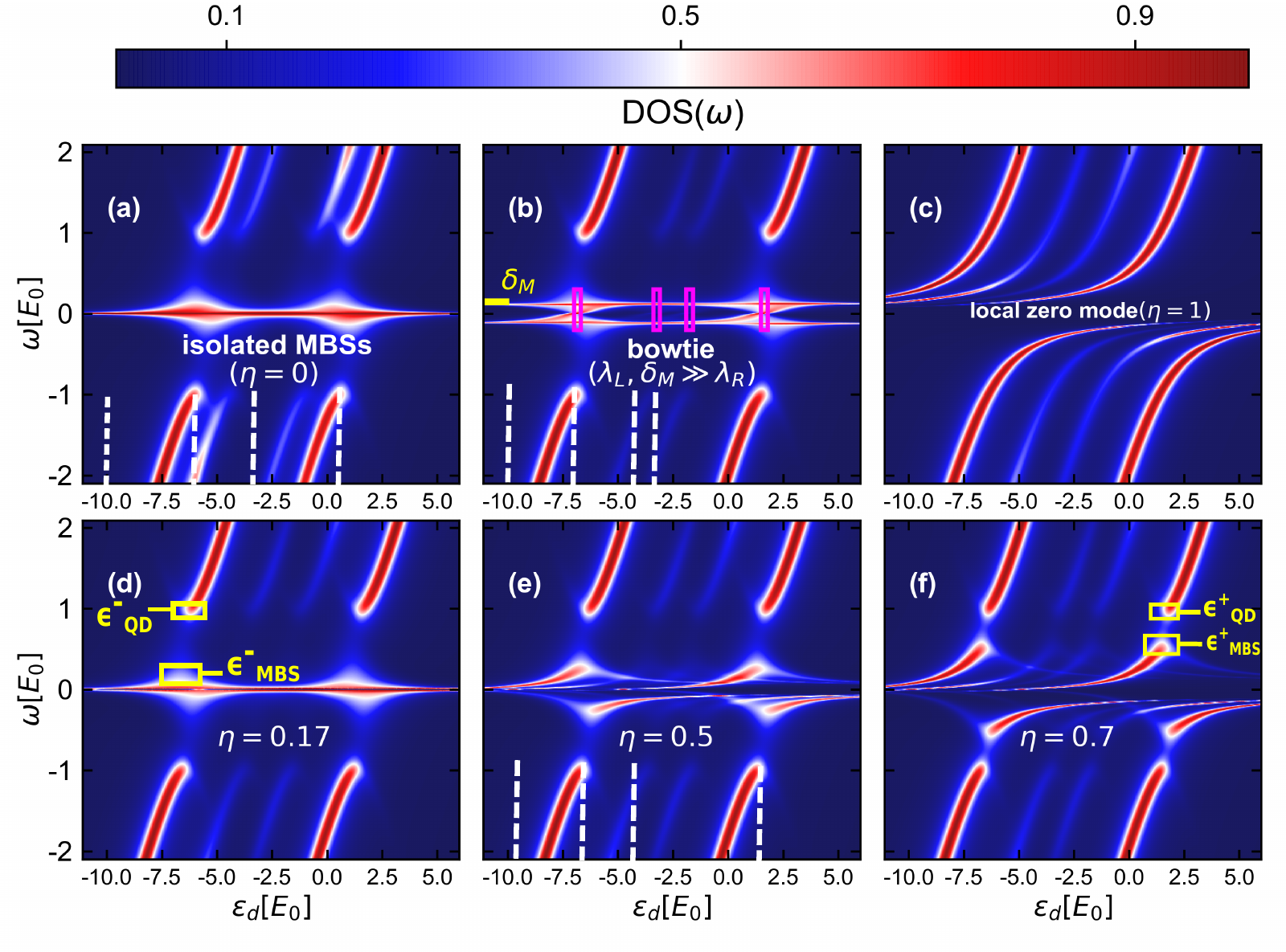}}
	\caption{Color scale plots of the DOS of a QD as a function of QD energy level $\varepsilon_{d}$ and spectral frequency $\omega$ for distinct regimes. In all situations we chose $U=5E_{0}$ and $\lambda_{L}=1.0E_{0}$. (a) highly nonlocal isolated MBSs ($\delta_{M}=\lambda_{R}=0$) for $V_{Z}=0.8E_{0}$ and $L=\SI{2.0}{\micro\meter}$; (b) bowtie profile, wherein $V_{Z}=1.72E_{0}$, $L=\SI{0.4}{\micro\meter}$, $\delta_{M}=0.12E_{0}$ and $\lambda_{R}=0.003E_{0}$. (c) regular fermionic zero mode ($\lambda_{L}=\lambda_{R}=E_{0}$, $\delta_{M}=0$), with $L=\SI{0.4}{\micro\meter}$ and $V_{Z}=1.38E_{0}$; (d)-(f) diamond profiles ($\delta_{M} \ll \lambda_{R},\lambda_{L}$), with actual nanowire length $L=\SI{0.4}{\micro\meter}$. Panel (d) exhibits the situation for  $V_{Z}=1.4E_{0}$, $\delta_{M}=0.004E_{0}$ and $\lambda_{R}=0.03E_{0}$. In (e) we have set $V_{Z}=1.5E_{0}$, $\delta_{M}=0.04E_{0}$ and $\lambda_{R}=0.25E_{0}$, while in panel (f) $V_{Z}=1.6E_{0}$, $\delta_{M}=0.08E_{0}$ and $\lambda_{R}=0.5E_{0}$. 
		\label{Fig2}}
\end{figure}

Fig.~\ref{Fig2}(a) corresponds to the highest nonlocal situation ($\eta=0$), where SC nanowire is long enough ($L\approx\num{2.0}\si{\micro\meter}$) to ensure formation of isolated MBSs at the ends ($\delta_{M}=0$). In this regime, the dot couples only with the left MBS ($\lambda_{R}=0$).  As predicted by the earlier works~\cite{RAguado,DJClarcke,AguadoExperiment}, the Majorana states remain unperturbed under variations of the QD energy level, since the latter can not cross the topologically protected zero-energy MBSs. 

For shorter wires ($L\approx\num{0.4}\si{\micro\meter}$), Fig.~\ref{Fig2}(b), DOS of the dot reveals the so-called \textit{``bowtie''} profile, characteristic to the regime when the overlap between MBSs is finite and the dot is only weakly hybridized with rightmost Majorana ($\lambda_{L},\delta_{M} \gg \lambda_{R}$)~\cite{RAguado}. In this situation the topological protection is absent and the energies of the overlapping MBSs are strongly perturbed in the vicinity of the resonance with the QD state. The splitting of near-zero states is ruled by direct hybridization between MBSs $2\delta_{M}$ (See yellow bar in the panel (b)). 

Fig.~\ref{Fig2}(c) demonstrates the spectra for the case of the  local fermionic zero-mode ($\delta_{M}=0$ and $\eta = 1$), corresponding to the highest localization of MBSs (lowest topological quality factor~\cite{DJClarcke}), for which any pronounced structure at $\omega=0$ is absent. 

The panels (d)-(f) of the Fig.~{\ref{Fig2}} correspond to the case of the shorter SC nanowires ($L\approx\num{0.4}\si{\micro\meter}$), for which $\delta_M\neq0$ but $\delta_{M} \ll \lambda_{R}$.  This regime corresponds to the situation wherein the wave function describing the right MBS moves towards the QD due to the application of the magnetic field \cite{Quantized,psABS2}.
One can notice the presence of the previously reported \textit{``diamond''} profiles~\cite{RAguado,AguadoExperiment}.  Fig.~{\ref{Fig2}}(d) shows the diamond lineshape for a quasi-ideal case of the isolated MBSs ($\eta = 0.17$), while panels (e) and (f) illustrate the situations where the nonlocal feature was suppressed by enhancing $\lambda_{R}$ and, consequently $\eta$. The loss of the nonlocality ($\eta \rightarrow1$) is related to the displacement of the right MBS ($\gamma_{R}$) wave function towards the left MBS, increasing the overlap between such states and enhancing the hybridization $\lambda_{R}$ of right Majorana mode with the dot state. By comparing the Figs.~{\ref{Fig2}}(d)-(f), it can be noticed that ABS formation due to the strong localization of the right MBS near the QD gives rise to the disintegration of the diamond shapes. In other words, the closer $\epsilon_{\text{MBS}}^{\pm}$ is to $\epsilon_{\text{QD}}^{\pm}$ (See panels (d) and (f)), the higher is the local nature of MBSs and the lower is the topological quality of the device.

\begin{figure}[t]
	\centerline{\includegraphics[width=3.6in,keepaspectratio]{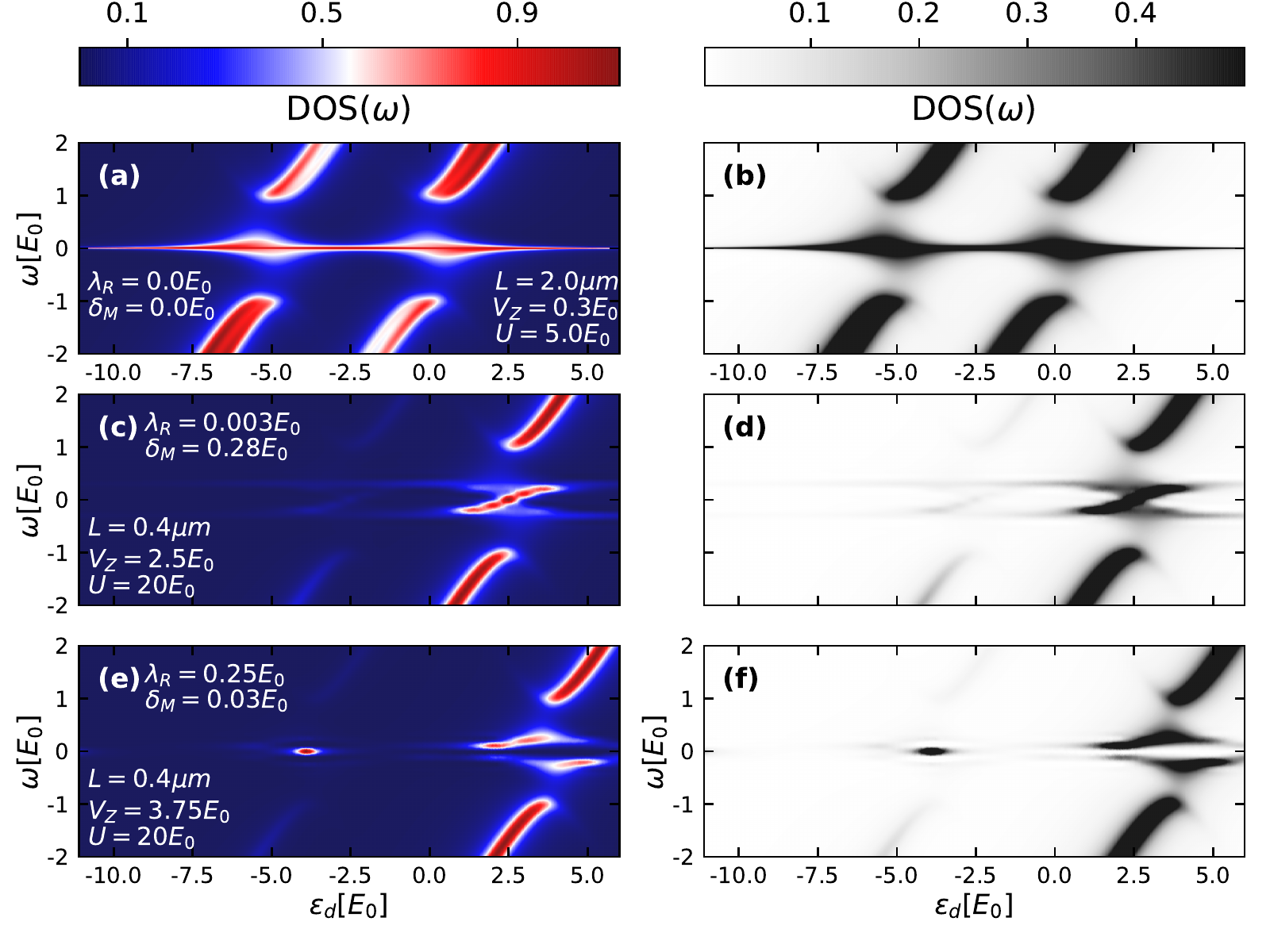}}
	\caption{Color scale plots of the DOS of a QD as a function of QD energy level $\varepsilon_{d}$ and spectral frequency $\omega$ for distinct regimes. In all situations we chose $\lambda_{L}=1.0E_{0}$. Left panels show (a)isolated MBSs, (c)bowtie and (e)diamond profiles, corresponding to the cases illusrtated by Fig.~\ref{Fig2}(a), (b) and (e), respectively, but for the distinct values of Zeeman splitting $V_{Z}$ and charging energy $U$. Right panels reproduce the corresponding left ones in the reduced gray scale. 
		\label{Newcolormap}}
\end{figure}

It is worth noting that the results presented in the Fig.~\ref{Fig2} differ from analogous results of the earlier works~\cite{RAguado,AguadoExperiment} due to the presence of extra crossing points appearing in the middle region of all the panels. This feature is direct outcome of the theoretical treatment afforded by Hubbard-I approximation~\cite{HubbardI} to charging energy term of the system Hamiltonian [Eq.~(\ref{eq:FullH})]. It is known that such approximation makes the condition of the transition to ferromagnetic state more restrictive compared to Hartree-Fock mean-field approximation, since it accounts for the higher order correlations thus reducing the energy of non-magnetic states with respect to ferromagnetic ones~\cite{HubbardI}. For this reason, Hartree-Fock mean-field approximation works well for larger Zeeman fields, as previously noticed by Prada \textit{et al}~\cite{RAguado}. However, within such a mean-field approach for any value of $V_{Z}$, the information about correlated motion of electrons is only taken into account with the mean occupation~\cite{Bruus}. Hubbard-I decoupling scheme accounts for such correlated motion, which gives rise to the appearance of the so-called Hubbard peaks at $\varepsilon_{d\sigma}$ and $\varepsilon_{d\sigma} + U$, describing the regime of the Coulomb blockade. From the experimental perspective, the work of Deng \textit{et al}~\cite{AguadoExperiment} shows the validity of the Hartree-Fock approach, since such an experiment was performed under relatively larger Zeeman fields, which are enough to resolve the QD spin-degrees of freedom. In this work, by using the Hubbard I decoupling scheme, we predict the system behavior in the scenario of weaker Zeeman splitting, which could be addressed in future experiments. 

Detuning of the parameters $V_{Z}$ and $U$ from the value corresponding to the Figs.~\ref{Fig2}(a), (b) and (e) allows us to recover low-energy spectrum theoretically predicted within the Hartree-Fock mean-field approximation~\cite{RAguado}, which was experimentally verified by Deng \textit{et. al}~\cite{AguadoExperiment}, thus ensuring the comprehensiveness of the approximation employed here. Figs.~\ref{Newcolormap}(a)-(b) illustrate the case of isolated MBSs, but with $V_{Z}$ lower than that of Fig.~\ref{Fig2}(a). Such a Zeeman splitting is unable to resolve spin up and down states of the QD. Consequently, instead of a four-peak structure, profiles with only two peaks resembling those presented in the Fig.~4(d) of the Ref.~\onlinecite{AguadoExperiment} appear. Figs.~\ref{Newcolormap}(c)-(d) show bowtie profiles corresponding to $V_{Z}=2.5E_{0}$ and $U=20E_{0}$ and comparable to those presented in the Fig.~3(b) of Ref.~\onlinecite{AguadoExperiment}. In the corresponding scenario only two peaks are present as well, since higher charging energy sets other peaks outside the considered range of $\varepsilon_{d}$. Increase of the value of $U$ results in the profiles shown in the Figs.~\ref{Newcolormap}(e)-(f) corresponding to the diamond situation shown in the Fig.~\ref{Fig2}(e) and comparable to those presented in the Fig.~3(c) of the Ref.~\onlinecite{AguadoExperiment}.

\begin{figure}[t]
	\centerline{\includegraphics[width=3.5in,keepaspectratio]{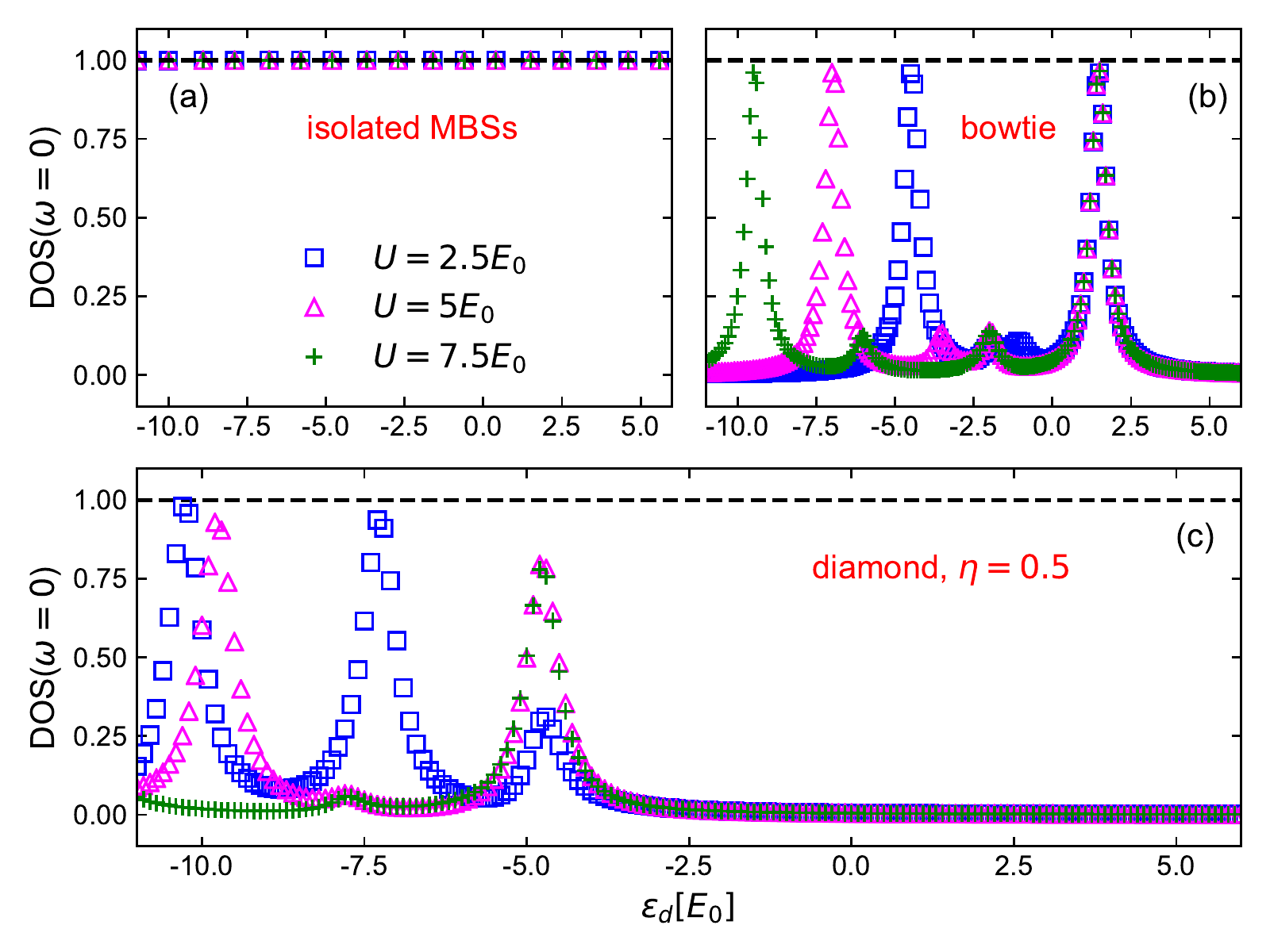}}
	\caption{ Density of states as a function of the QD level $\varepsilon_{d}$ at $\omega=0$ for (a) isolated MBSs, (b) bowtie and (c) diamond configurations. Plots for several  values of charging energy [$U=2.5E_{0}$ (blue squares), $5E_{0}$ (magenta triangles) and $7.5E_{0}$ (green crosses)] are presented.
		\label{Fig3}}
\end{figure}

\begin{figure}[t]
	\centerline{\includegraphics[width=3.5in,keepaspectratio]{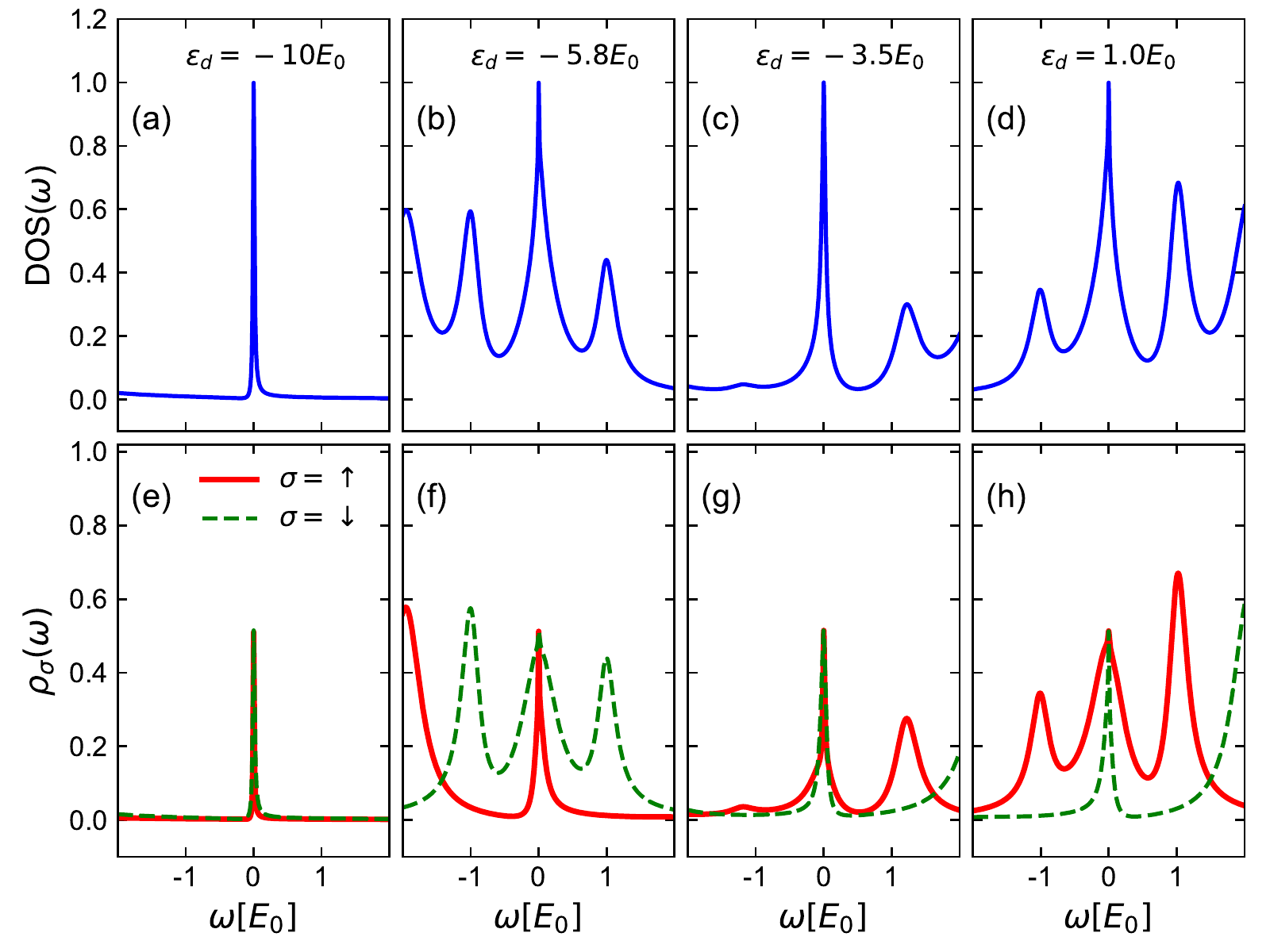}}
	\caption{(a)-(d): (a)-(d): total DOS as a function of $\omega$ for MBSs (corresponding to the Fig.2(a)) for various values of $\varepsilon_{d}$ corresponding to the vertical dashed white lines in Fig.~\ref{Fig2}(a). (e)-(h) spin resolved DOS $\rho_{\sigma}(\omega)$ for same parameters as in the panels (a)-(d)
		\label{Isolated}}
\end{figure}

\begin{figure}[t]
	\centerline{\includegraphics[width=3.5in,keepaspectratio]{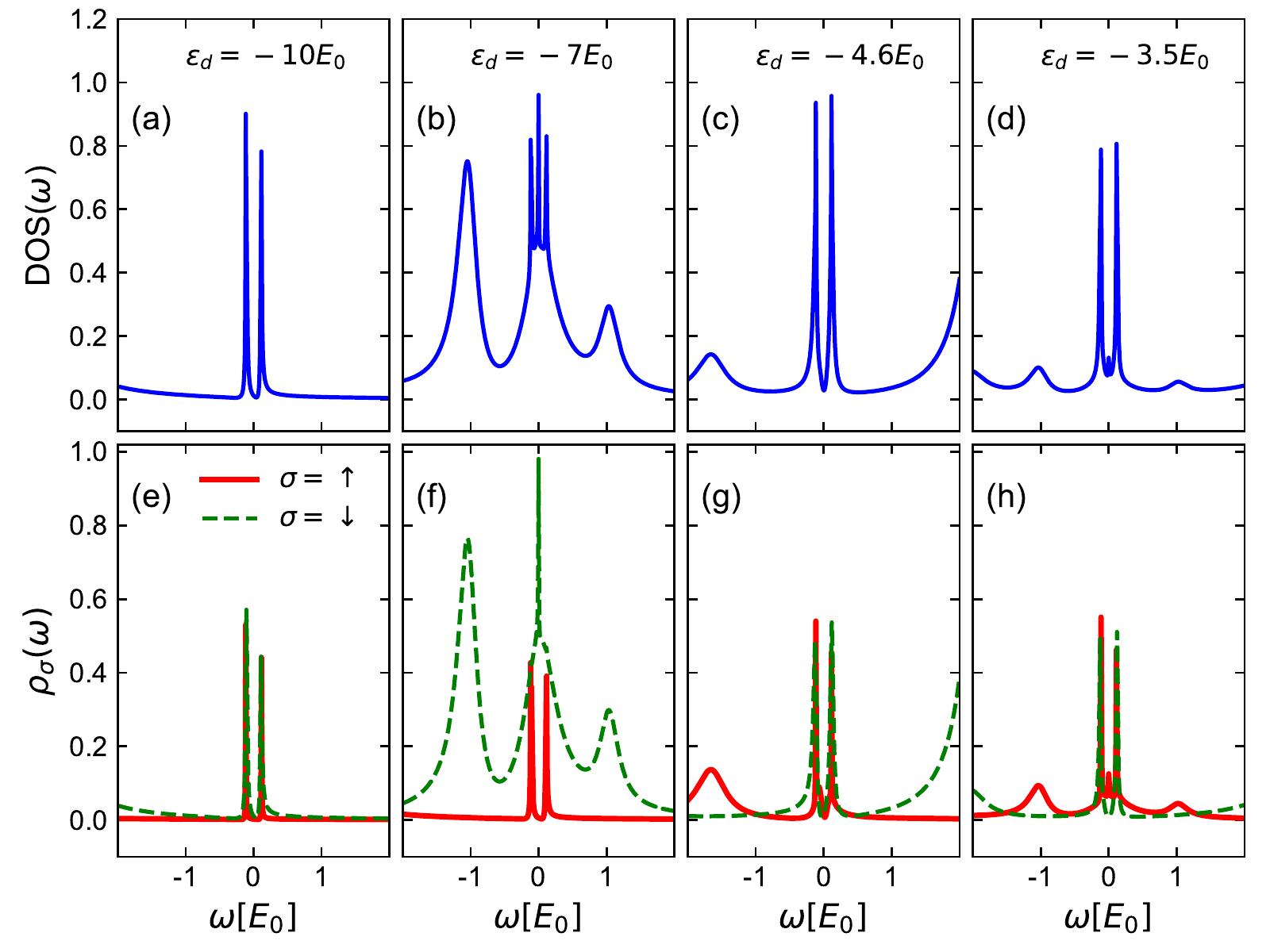}}
	\caption{(a)-(d): total DOS as a function of $\omega$ for bowtie profile (corresponding to the Fig.2(b)) for various values of $\varepsilon_{d}$ corresponding to the vertical dashed white lines in Fig.~\ref{Fig2}(b). (e)-(h) spin resolved DOS $\rho_{\sigma}(\omega)$ for same parameters as in the panels (a)-(d)
		\label{Bowtie}}
\end{figure}

Fig~\ref{Fig3} shows the DOS of the dot as a function of QD level $\varepsilon_{d}$ at $\omega=0$ for (a) isolated MBSs, (b) bowtie and (c) diamond situations, for distinct strengths of the dot charging energy [$U=2.5E_{0}$ (blue squares), $5E_{0}$ (magenta triangles) and $7.5E_{0}$ (green crosses) ]. In the highest nonlocal case [Fig~\ref{Fig3}(a)], the insensitivity of the zero frequency peak to the tuning of the QD level and variations of the dot charging energy is verified~\cite{RAguado,DJClarcke,AguadoExperiment}. There is a plateau in the total DOS characteristic to ZBCP. This scenario breaks down for the  possible situation of the formation of ABS due to the overlap between MBSs [Fig~\ref{Fig3}(b-c)]. In this case the plateau in the DOS is destroyed and positions of the peaks change with variations of both position of the dot level $\varepsilon_{d}$ and charging energy $U$. Fig~\ref{Fig3}(b) describes a linecut at $\omega=0$ for the bowtie configuration, wherein the four resonances for $U=5E_{0}$ (magenta triangles) correspond to the anticrossings between the dot level and near zero-energy states appearing in the Fig~\ref{Fig2}(b) (indicated by magenta rectangles). These anticrossing points (resonance positions) are strongly dependent on the charging energy, since near zero-energy ABS, which can be a trivial non-protected state, is affected by the QD energy levels. Similar behavior is found for a diamond profile with degree of nonlocality $\eta =0.5$, as it is shown in the Fig.~\ref{Fig3}(c). Moreover, the plateau depicted in the Fig.~\ref{Fig3}(a) also allows to distinguish the Majorana ZBCP from that induced by usual Kondo effect, since Kondo resonance only appears when a QD is single occupied, $\varepsilon_{d}<\varepsilon_{F}$, as verified by one of us in the Ref.~\onlinecite{Vernek} (see Fig.~1(g)-(i) of that paper).

\begin{figure}[t]
	\centerline{\includegraphics[width=3.50in,keepaspectratio]{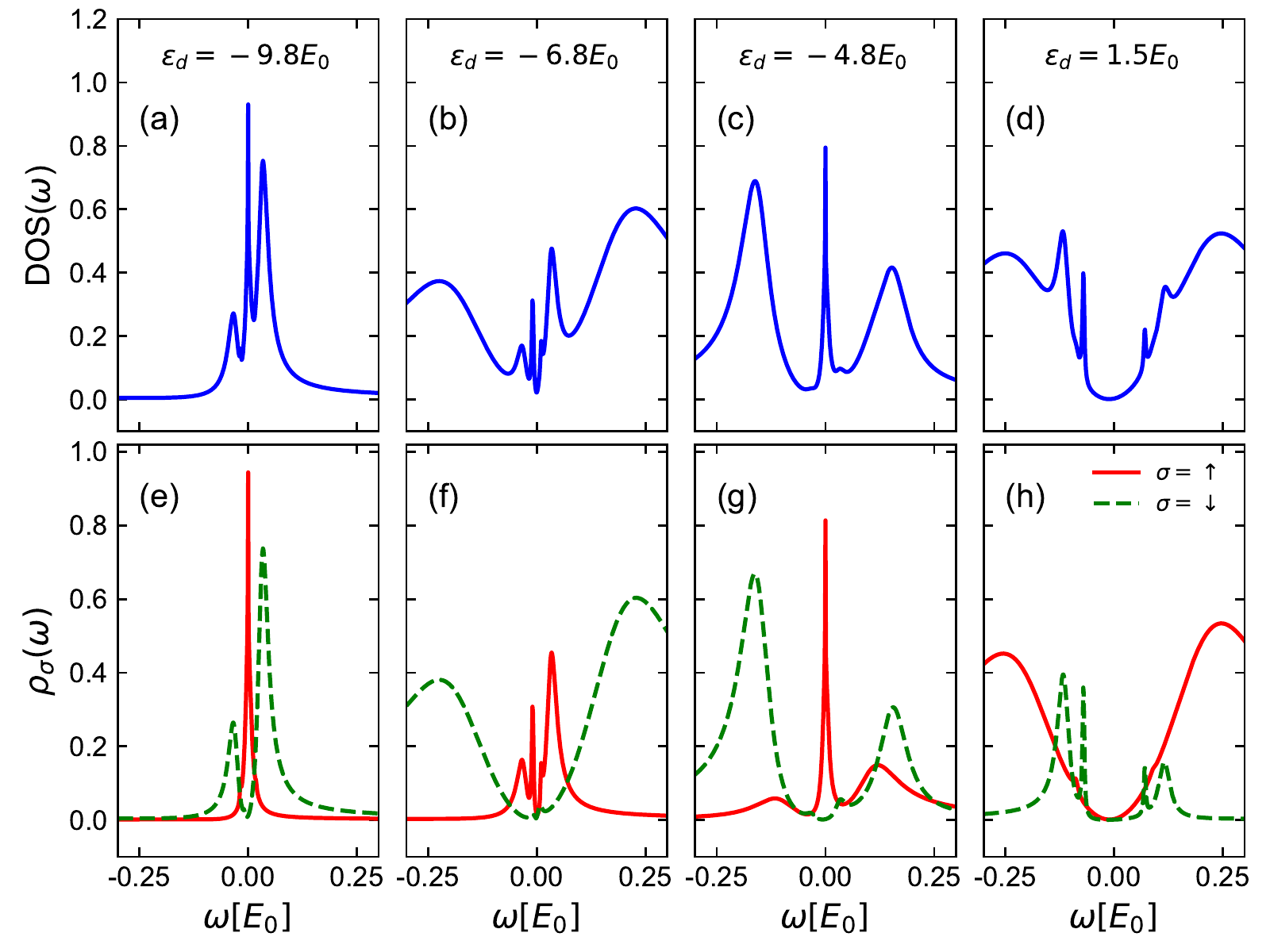}}
	\caption{(a)-(d): total DOS as a function of $\omega$ for diamond profile for various values of $\varepsilon_{d}$ corresponding to the vertical dashed white lines in Fig.~\ref{Fig2}(e). (e)-(h) spin resolved DOS $\rho_{\sigma}(\omega)$ for same parameters as in the panels (a)-(d)
		\label{Diamond}}
\end{figure}

The impact of each spin component to the total DOS is presented in the Figs.~\ref{Isolated}, \ref{Bowtie} and \ref{Diamond} for isolated MBSs, bowtie [Fig.\ref{Fig2}(b)] and diamond [Fig.\ref{Fig2}(e)] configurations, respectively. Panels (a)-(d) of the  Fig.~\ref{Isolated} show the DOS of the dot for isolated MBSs configuration [Fig.\ref{Fig2}(a)] as a function of $\omega$ for several values of $\varepsilon_{d}$ indicated by the white dashed lines in the Fig.~\ref{Fig2}(a). As can be seen, a peak at $\omega=0$ emerges for all values of the dot energy as observed in the panels (a)-(d). Panels (e)-(h) of the same figure reveal that zero-peak structure in the total DOS is spin degenerated: at $\omega=0$,  $\rho_{\uparrow}=\rho_{\downarrow}$. 

This degeneracy is broken when MBSs overlap ($\delta_{M}\neq0$) and the dot hybridizes with the rightmost Majorana ($\lambda_{R} \neq 0$) as well, as it is demonstrated for the bowtie configuration in the Fig.~\ref{Bowtie}. In this case a zero-peak structure in the  total DOS emerges only when $\varepsilon_{d}$ crosses zero-energy as it happens in the panel (b).  In this case DOS around $\omega=0$ becomes spin sensitive, $\rho_{\uparrow}\neq\rho_{\downarrow}$, as one can verify from the panel (f), thus suggesting the situation of the  spin-dependent transport. 

Diamond configuration also displays spin-dependent behavior, as it is shown in the Fig.~\ref{Diamond}. However, there is a remarkable difference from the bowtie case: the near zero-energy two-peaks having $\sigma=\uparrow$ are no longer pinned and merge with one another at $\omega=0$ [panels (e) and (g)] for certain values of $\varepsilon_{d}$, giving rise to zero-peak structure in the DOS, as displayed in the panels (a) and (c). Such a feature is consistent with the picture of coalescing ABSs (overlapped MBSs), which can mimic the behavior MBSs under certain conditions~\cite{ABS2,ABS3,ABS4,psABS1,psABS2}. It also can be verified for the situations where the QD level does not shift the near-zero peaks describing ABSs towards $\omega=0$ [panels(f) and (h)] and consequently there is no peak structure at zero-energy in DOS [panels (b) and (d)]. 

It is worth noting that the presence of ABSs in the total DOS of a QD [Eq.~(\ref{eq:DOS})] in the results presented in the Figs.~\ref{Fig3}, \ref{Bowtie} and \ref{Diamond} can be confirmed by the computation of the anomalous Green's function $G_{f_{\uparrow}^{\dagger},d_{\sigma}}^{r}(\omega)$. The latter enters into the Green's function of the QD according to the Eq.~(\ref{eq:A4}),  describes the correlation between  QD ($d^{\dagger}_{\sigma}$) and SC nanowire appearing due Andreev reflection (AR) ~\cite{ABSLee,Baraski2013,RevModPhysimpuritiesSC} and can be defined in the time domain as:~\cite{Bruus}

\begin{equation}
G_{f_{\uparrow}^{\dagger},d_{\sigma}}^{r}(t-t')=-\imath \theta (t-t')\langle \lbrace f_{\uparrow}^{\dagger}(t),d_{\sigma}^{\dagger} (t') \rbrace  \rangle, \label{eq:fdagd}  
\end{equation}wherein $\theta(t-t')$ is Heaviside function, $\lbrace...,...\rbrace$ denotes anticommutator and $f_{\uparrow}^{\dagger}$ corresponds to the creation of a non-local fermion in SC nanowire, which is formed by the linear combination of the left ($\gamma_{L}$) and right ($\gamma_{R}$) MBSs, \textit{i.e,} $f_{\uparrow}^{\dagger}=(\gamma_{L}-\imath\gamma_{R})/\sqrt{2}$ and $f_{\uparrow}=(\gamma_{L} + \imath\gamma_{R})/\sqrt{2}$.

\begin{figure}[t]
	\centerline{\includegraphics[width=3.5in,keepaspectratio]{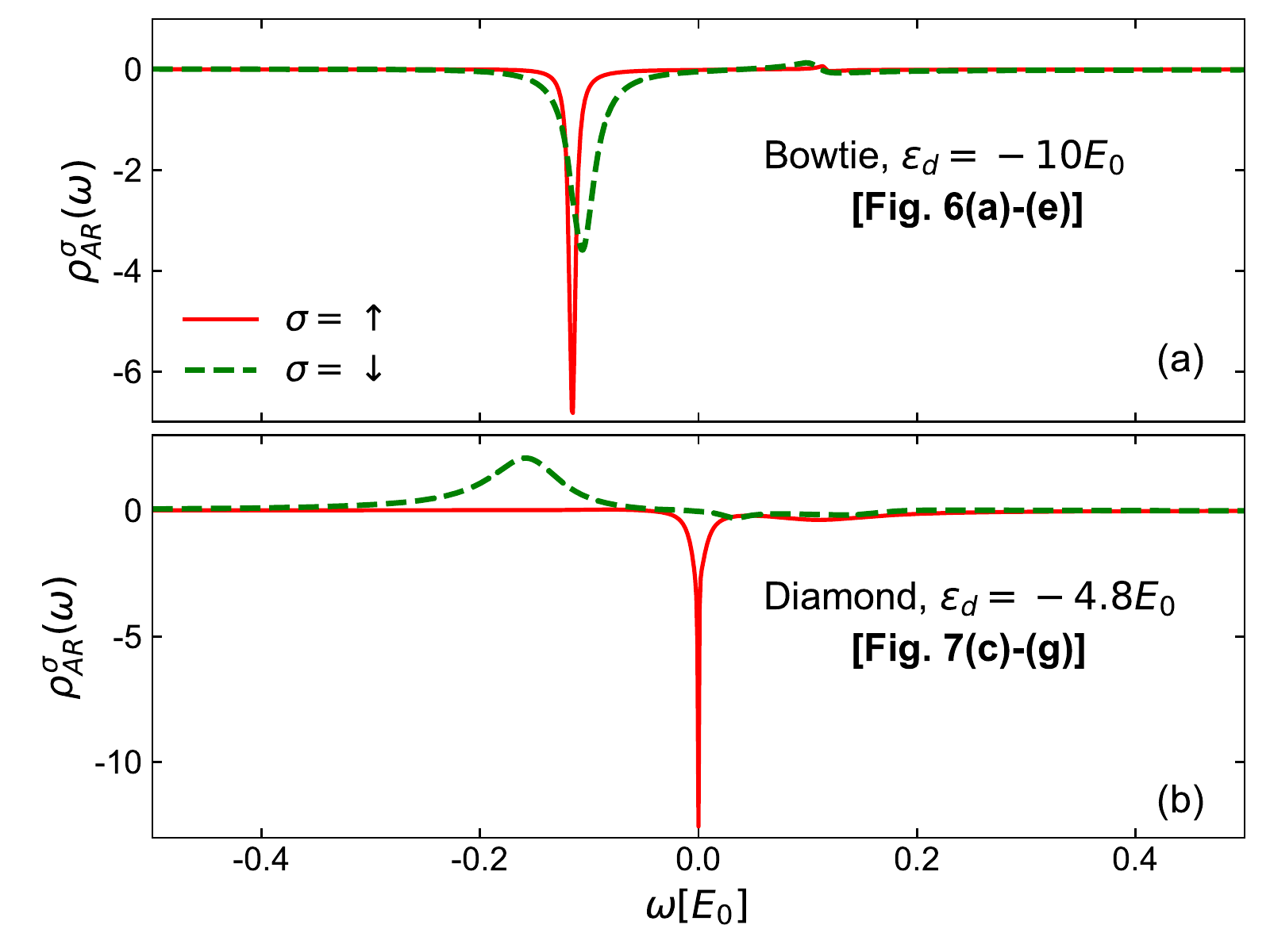}}
	\caption{DOS related to AR processes for each spin component, wherein two representative situations were considered. Panel (a) depicts the same case of Figs.~\ref{Bowtie}(a)-(e), while panel (b) is related to Figs.~\ref{Diamond}(c)-(g).} 
		\label{ARFanodips}
\end{figure}

AR can take place through different transport channels, once QD is coupled to the both ends of a nanowire ($\lambda_{L}$ and $\lambda_{R}\neq0$). The presence of the coupling asymmetry can give rise to the appearence of Fano antiresonances ~\cite{RevFano,CPS,Recher,MBSoscillations} as it can be seen in the Fig.~\ref{ARFanodips}, where we illustrate AR process by plotting $\rho_{AR}^{\sigma}(\omega) = -\text{Im} \lbrace G_{f_{\uparrow}^{\dagger},d_{\sigma}}^{r}(\omega)\rbrace/\pi$ for two representative \textit{``bowtie''} and \textit{``diamond''} configurations (see Figs.~\ref{Bowtie}(a)-(e) and~\ref{Diamond}(c)-(g), respectively).

\section{Conclusions}

We presented the theoretical study of MBSs nonlocality in the hybrid device sketched in the Fig.~\ref{Setup}. It was analyzed in detail how the charging energy affects spin resolved DOS of a QD coupled with MBSs. For highly nonlocal MBSs there is a plateau at zero-energy in the QD density of states for any values of the dot level and charging energy. For overlapping MBSs the spectrum of the dot is strongly modified.  It was shown that the zero-peak structure in the DOS reveals pronounced spin dependence if MBSs become hybridized. Our findings suggest that a spin-dependent local probe may be used as a tool to resolve an outstanding problem in experimental Majorana physics: discriminating between the cases when ZBCP is due to the isolated MBSs or ABSs.
\section{Acknowledgments} 

We thank the funding Brazilian agencies CNPq (Grants No.~307573/2015-0 and No.~305668/2018-8) and S\~ao Paulo Research Foundation (FAPESP) Grant No.~2015/23539-8. IAS acknowledges support by Horizon2020 RISE project CoExAN. MdS acknowledges the Austrian Academy of Science \"OAW for the JESH fellowship and Serdar Sariciftci for the hospitality.
\appendix
\section{Quantum dot Green's function derivation}\label{AppendixA}

In this Appendix we present the main steps concerning on derivation of Eq.~(\ref{eq:E1}) via EOM technique. It is known that such a method can be summarized according to~\cite{Jauho,Bruus}
\begin{equation}
(\omega+\imath 0^{+})G_{c_{i}c_{j}}^{r}(\omega)=\{c_{i},c_{j}^{\dagger}\}+G_{[c_{i},\mathcal{H}];c_{j}}^{r}(\omega),\label{eq:EOM}
\end{equation}
wherein $c_{i(j)}$ is a fermionic operator belonging to Hamiltonian $\mathcal{H}$. Hence, the Green's function of QD is given by:
\begin{equation}
(\omega+\imath 0^{+})G_{d_{\sigma}d_{\sigma}}^{r}(\omega) = 1+G_{[d_{\sigma},H];d_{\sigma}}^{r},
\end{equation}
with
\begin{eqnarray}
[d_{\sigma},H] & = & \varepsilon_{d\sigma}d_{\sigma} + Ud_{\sigma}n_{d\bar{\sigma}} + \sqrt{2}V\sum_{\mathbf{k}}e_{\mathbf{k}\sigma} \nonumber \\ 
& - & \lambda \sum_{\tilde{\sigma}}\delta_{\sigma\tilde{\sigma}}f_{\uparrow} - \lambda' \sum_{\tilde{\sigma}}\delta_{\sigma\tilde{\sigma}}f_{\uparrow}^{\dagger}\label{eq:A3},
\end{eqnarray}
where $f_{\uparrow}^{\dagger}(f_{\uparrow})$ stands for creation(annihilation) of a nonlocal fermion in the hybrid SC nanowire, since it can be rewritten as a linear combination of MBSs at opposite ends of the nanowire. Within this picture, the coupling strengths between the QD and the nonlocal fermionic site are given by $\lambda = (|\lambda_{L}|-|\lambda_{R}|)/\sqrt{2}$ and $\lambda' = (|\lambda_{L}|+|\lambda_{R}|)/\sqrt{2}$, respectively. Eq~(\ref{eq:A3}) allows to write that
\begin{align}
(\omega^{+}-\varepsilon_{d\sigma})G_{d_{\sigma}d_{\sigma}}^{r}(\omega)=\nonumber\\
1+UG_{d_{\sigma}n_{d\bar{\sigma}},d_{\sigma}}^{r}(\omega) & +\sqrt{2}V\sum_{\mathbf{k}}G_{e_{\mathbf{k}\sigma}d_{\sigma}}^{r}(\omega)\nonumber\\
-\lambda\sum_{\tilde{\sigma}}\delta_{\sigma\tilde{\sigma}}G_{f_{\uparrow},d_{\sigma}}^{r}(\omega) & -\lambda'\sum_{\tilde{\sigma}}\delta_{\sigma\tilde{\sigma}} G_{f_{\uparrow}^{\dagger},d_{\sigma}}^{r}(\omega), \label{eq:A4}
\end{align}
with $\omega^{+}\rightarrow \omega + \imath 0^{+}$. The three last Green's functions also are obtained through straightforward application of EOM technique [Eq.~(\ref{eq:EOM})], being respectively given by:
\begin{equation}
G_{e_{\mathbf{k}\sigma}d_{\sigma}}^{r}(\omega)=\sqrt{2}V\sum_{\mathbf{k}}\frac{1}{(\omega^{+} -\varepsilon_{\mathbf{k}\sigma} )}G_{d_{\sigma}d_{\sigma}}^{r}(\omega),\label{eq:A5}
\end{equation}
\begin{align}
(\omega^{+}-\delta_{\text{M}})G_{f_{\uparrow}d_{\sigma}}^{r}(\omega)=\nonumber\\-\lambda\sum_{\tilde{\sigma}}G_{d_{\tilde{\sigma}}d_{\sigma}}^{r}(\omega) & +\lambda'\sum_{\tilde{\sigma}}G_{d_{\tilde{\sigma}}^{\dagger}d_{\sigma}}^{r}(\omega)\label{eq:A6}
\end{align}
and
\begin{align}
(\omega^{+}+\delta_{\text{M}})G_{f_{\uparrow}^{\dagger}d_{\sigma}}^{r}(\omega)=\nonumber\\ \lambda \sum_{\tilde{\sigma}} G_{d_{\tilde{\sigma}}^{\dagger}d_{\sigma}}^{r}(\omega) & -\lambda' \sum_{\tilde{\sigma}} G_{d_{\tilde{\sigma}}d_{\sigma}}^{r}(\omega). \label{eq:A7}
\end{align}
Substituting Eqs.~(\ref{eq:A5}), (\ref{eq:A6}) and (\ref{eq:A7}) into Eq.~(\ref{eq:A4}), we find 
\begin{align}
\left(\omega^{+}-\varepsilon_{d\sigma}+\imath\Gamma-\frac{\lambda{}^{2}}{\omega^{+}-\delta_{\text{M}}}-\frac{\lambda'{}^{2}}{\omega^{+}+\delta_{\text{M}}}\right)G_{d_{\sigma}d_{\sigma}}^{r}(\omega)=\nonumber\\
1+UG_{d_{\sigma}n_{d\bar{\sigma}},d_{\sigma}}^{r}(\omega)-(2\lambda\lambda')KG_{d_{\sigma}^{\dagger}d_{\sigma}}^{r}(\omega), \label{eq:A8}
\end{align}
where $K$ is given by Eq.~(\ref{eq:K}) and $\Gamma$ is the Anderson parameter~\cite{Anderson}. We now evaluate the Green's function $G_{d_{\sigma}^{\dagger}d_{\sigma}}^{r}(\omega)$, getting the following result:
\begin{align}
\left(\omega^{+}+\varepsilon_{d\sigma}+\imath\Gamma-\frac{\lambda{}^{2}}{\omega^{+}+\delta_{\text{M}}}-\frac{\lambda'^{2}}{\omega^{+}-\delta_{\text{M}}}\right)G_{d_{\sigma}^{\dagger}d_{\sigma}}^{r}(\omega)=\nonumber\\
-UG_{d_{\sigma}^{\dagger}n_{d\bar{\sigma}}d_{\sigma}}^{r}(\omega)-(2\lambda\lambda')KG_{d_{\sigma}d_{\sigma}}^{r}(\omega).\label{eq:A9}
\end{align}
Substituting into Eq.~(\ref{eq:A8}) and recognizing $\bar{K}_{\sigma}$ [Eq.~(\ref{eq:Kbar})], $K_{1}$[Eq.~(\ref{eq:K1})] and $K_{2}$[Eq.~(\ref{eq:K2})], is now easy to show that Eq.~(\ref{eq:A8}) becomes into Eq.~(\ref{eq:E1}). 

\section{Hubbard-I Approximation} \label{AppendixB}

We now evaluate the two particle Green's functions of Eq.~(\ref{eq:E1}) according to EOM. Following Eq.~(\ref{eq:EOM}), we have
\begin{equation}
(\omega+\imath\eta^{+})G_{d_{\sigma}n_{d\bar{\sigma}},d_{\sigma}}^{r}(\omega)=\langle n_{d\bar{\sigma}}\rangle+G_{[d_{\sigma}n_{d\bar{\sigma}},H];d_{\sigma}}^{r}(\omega). \label{eq:B1}
\end{equation}
Deriving the commutator
\begin{align}
[d_{\sigma}n_{d\bar{\sigma}},H] = \varepsilon_{d\sigma}d_{\sigma}n_{d\bar{\sigma}} & + Ud_{\sigma}n_{d\bar{\sigma}} \nonumber\\
+\sqrt{2}V\sum_{\mathbf{k}}(-e_{\mathbf{k}\bar{\sigma}}^{\dagger}d_{\bar{\sigma}}d_{\sigma}+&d_{\bar{\sigma}}^{\dagger}e_{\mathbf{k}\bar{\sigma}}d_{\sigma}+e_{\mathbf{k}\sigma}n_{d\bar{\sigma}}) \nonumber \\
+\lambda\sum_{\tilde{\sigma}}(-\delta_{\tilde{\sigma}\bar{\sigma}}d_{\bar{\sigma}}f_{\uparrow}^{\dagger}d_{\sigma}+&\delta_{\tilde{\sigma}\bar{\sigma}}f_{\uparrow}d_{\bar{\sigma}}^{\dagger}d_{\sigma}-\delta_{\tilde{\sigma}\sigma}f_{\uparrow}n_{d\bar{\sigma}})\nonumber \\
+ \lambda'\sum_{\tilde{\sigma}}(-\delta_{\tilde{\sigma}\bar{\sigma}}d_{\bar{\sigma}}f_{\uparrow}d_{\sigma}+&\delta_{\tilde{\sigma}\bar{\sigma}}f_{\uparrow}^{\dagger}d_{\bar{\sigma}}^{\dagger}d_{\sigma}-\delta_{\tilde{\sigma}\sigma}f_{\uparrow}^{\dagger}n_{d\bar{\sigma}}), \label{eq:B2}
\end{align}
we find
\begin{align}
(\omega^{+}-\varepsilon_{d\sigma}-U)G_{d_{\sigma}n_{d\bar{\sigma}},d_{\sigma}}^{r}(\omega) &= \langle n_{d\bar{\sigma}}\rangle   \nonumber\\ +\sqrt{2}V\sum_{\mathbf{k}}G_{e_{\mathbf{k}\sigma}n_{d\bar{\sigma}},d_{\sigma}}^{r}(\omega) & +\sqrt{2}V\sum_{\mathbf{k}} G_{d_{\bar{\sigma}}^{\dagger}e_{\mathbf{k}\bar{\sigma}}d_{\sigma}d_{\sigma}}^{r}(\omega) \nonumber \\
-\sqrt{2}V\sum_{\mathbf{k}}G_{e_{\mathbf{k}\bar{\sigma}}^{\dagger}d_{\bar{\sigma}}d_{\sigma},d_{\sigma}}^{r}(\omega) & -\lambda\sum_{\tilde{\sigma}}\delta_{\tilde{\sigma}\sigma}G_{f_{\uparrow}n_{d\bar{\sigma}}d_{\sigma}}^{r}(\omega)\nonumber \\
+ \lambda\sum_{\tilde{\sigma}}\delta_{\tilde{\sigma}\bar{\sigma}}G_{f_{\uparrow}d_{\bar{\sigma}}^{\dagger}d_{\sigma}d_{\sigma}}^{r}(\omega)&-\lambda\sum_{\tilde{\sigma}}\delta_{\tilde{\sigma}\bar{\sigma}}G_{d_{\bar{\sigma}}f_{\uparrow}^{\dagger}d_{\sigma}d_{\sigma}}^{r}(\omega) \nonumber \\
-\lambda' \sum_{\tilde{\sigma}}\delta_{\tilde{\sigma}\sigma}G_{f_{\uparrow}^{\dagger}n_{d\bar{\sigma}}d_{\sigma}}^{r}(\omega) & + \lambda'\sum_{\tilde{\sigma}}\delta_{\tilde{\sigma}\bar{\sigma}}G_{f_{\uparrow}^{\dagger}d_{\bar{\sigma}}^{\dagger}d_{\sigma};d_{\sigma}}^{r}(\omega)\nonumber \\
-\lambda' \sum_{\tilde{\sigma}}\delta_{\tilde{\sigma}\bar{\sigma}}G_{d_{\bar{\sigma}}f_{\uparrow}d_{\sigma}d_{\sigma}}^{r}(\omega). & \label{eq:B3}
\end{align}
At this point we apply the Hubbard-I decoupling scheme~\cite{HubbardI} by considering the following approximations:
\begin{equation}
G_{d_{\bar{\sigma}}^{\dagger}e_{\mathbf{k}\bar{\sigma}}d_{\sigma}d_{\sigma}}^{r}(\omega)\approx\langle d_{\bar{\sigma}}^{\dagger}e_{\mathbf{k}\bar{\sigma}}\rangle G_{d_{\sigma}d_{\sigma}}^{r}(\omega),
\end{equation}
\begin{equation}
G_{e_{\mathbf{k}\bar{\sigma}}^{\dagger}d_{\bar{\sigma}}d_{\sigma},d_{\sigma}}^{r}(\omega)\approx\langle e_{\mathbf{k}\bar{\sigma}}^{\dagger}d_{\bar{\sigma}}\rangle G_{d_{\sigma}d_{\sigma}}^{r}(\omega),
\end{equation}
\begin{equation}
G_{a_{\uparrow}d_{\bar{\sigma}}^{\dagger}d_{\sigma}d_{\sigma}}^{r}(\omega)\approx\langle a_{\uparrow}d_{\bar{\sigma}}^{\dagger}\rangle G_{d_{\sigma}d_{\sigma}}^{r}(\omega),
\end{equation}
\begin{equation}
G_{d_{\bar{\sigma}}a_{\uparrow}^{\dagger}d_{\sigma}d_{\sigma}}^{r}(\omega)\approx\langle d_{\bar{\sigma}}a_{\uparrow}^{\dagger}\rangle G_{d_{\sigma}d_{\sigma}}^{r}(\omega),
\end{equation}
\begin{equation}
G_{a_{\uparrow}^{\dagger}d_{\bar{\sigma}}^{\dagger}d_{\sigma};d_{\sigma}}^{r}(\omega)\approx\langle a_{\uparrow}^{\dagger}d_{\bar{\sigma}}^{\dagger}\rangle G_{d_{\sigma}d_{\sigma}}^{r}(\omega),
\end{equation}
and
\begin{equation}
G_{d_{\bar{\sigma}}a_{\uparrow}d_{\sigma}d_{\sigma}}^{r}(\omega)\approx\langle d_{\bar{\sigma}}a_{\uparrow}\rangle G_{d_{\sigma}d_{\sigma}}^{r}(\omega).
\end{equation}
Taking into account that $\langle d_{\bar{\sigma}}^{\dagger}e_{\mathbf{k}\bar{\sigma}}\rangle=\langle e_{\mathbf{k}\bar{\sigma}}^{\dagger}d_{\bar{\sigma}}\rangle$, $\langle a_{\uparrow}d_{\bar{\sigma}}^{\dagger}\rangle=\langle d_{\bar{\sigma}}a_{\uparrow}^{\dagger}\rangle$ and $\langle a_{\uparrow}^{\dagger}d_{\bar{\sigma}}^{\dagger}\rangle=\langle d_{\bar{\sigma}}a_{\uparrow}\rangle$, Eq.~(\ref{eq:B3}) becomes into 
\begin{align}
(\omega^{+}-\varepsilon_{d\sigma}-U)G_{d_{\sigma}n_{d\bar{\sigma}},d_{\sigma}}^{r}(\omega) & =\langle n_{d\bar{\sigma}}\rangle\nonumber\\
\sqrt{2}V\sum_{\mathbf{k}}G_{e_{\mathbf{k}\sigma}n_{d\bar{\sigma}},d_{\sigma}}^{r}(\omega) & -\lambda\sum_{\tilde{\sigma}}\delta_{\tilde{\sigma}\sigma}G_{f_{\uparrow}n_{d\bar{\sigma}}d_{\sigma}}^{r}(\omega)\nonumber\\
-  \lambda'\sum_{\tilde{\sigma}}\delta_{\tilde{\sigma}\sigma}G_{f_{\uparrow}^{\dagger}n_{d\bar{\sigma}}d_{\sigma}}^{r}(\omega).\label{eq:B10}
\end{align}
As can be seen in the procedure above, we threw away the Green's functions which describe spin-flip mechanisms between the QD level and the metallic lead, thus leading to the impossibility of catching Kondo-type correlations~\cite{Bruus,Jauho}. 

The other two-particle Green's functions of equation above also are found with the EOM, followed by approximations introduced by Hubbard-I procedure. Hence, we get
\begin{equation}
G_{e_{\mathbf{k}\sigma}n_{d\bar{\sigma}},d_{\sigma}}^{r}(\omega)  =  \sqrt{2}V\sum_{\mathbf{k}}\frac{1}{(\omega^{+}-\varepsilon_{\mathbf{k}\sigma})}G_{d_{\sigma}n_{d\bar{\sigma}},d_{\sigma}}^{r}(\omega), \label{eq:B11}
\end{equation}
\begin{align}
(\omega^{+}-\delta_{\text{M}})\sum_{\tilde{\sigma}}\delta_{\tilde{\sigma}\sigma}G_{f_{\uparrow}n_{d\bar{\sigma}},d_{\sigma}}^{r}(\omega) = & \nonumber \\
-\lambda G_{d_{\sigma}n_{d\bar{\sigma}},d_{\sigma}}^{r}(\omega)& +\lambda' G_{d_{\sigma}^{\dagger}n_{d\bar{\sigma}},d_{\sigma}}^{r}(\omega) \label{eq:B12}
\end{align}
and
\begin{align}
(\omega^{+}+\delta_{\text{M}})\sum_{\tilde{\sigma}}\delta_{\tilde{\sigma}\sigma}G_{f_{\uparrow}^{\dagger}n_{d\bar{\sigma}},d_{\sigma}}^{r}(\omega) = \nonumber \\
\lambda G_{d_{\sigma}^{\dagger}n_{d\bar{\sigma}},d_{\sigma}}^{r}(\omega)&-\lambda' G_{d_{\sigma}n_{d\bar{\sigma}},d_{\sigma}}^{r}(\omega)\label{eq:B13}, 
\end{align}
allowing us to find that 
\begin{align}
(\omega^{+}-\varepsilon_{d\sigma}-U-K1 + \imath\Gamma)G_{d_{\sigma}n_{d\bar{\sigma}},d_{\sigma}}^{r}(\omega) & = \langle n_{d\bar{\sigma}}\rangle\nonumber \\
- (2\lambda\lambda') KG_{d_{\sigma}^{\dagger}n_{d\bar{\sigma}},d_{\sigma}}^{r}(\omega).& \label{eq:B14}
\end{align}
By adopting the same procedure described above, we evaluate $G_{d_{\sigma}^{\dagger}n_{d\bar{\sigma}},d_{\sigma}}^{r}(\omega)$, resulting in Eq.~(\ref{eq:E3}), which allows us to rewrite Eq.~(\ref{eq:B14}) as expressed in Eq.~(\ref{eq:E2}). It is worth noting that turning off the QD-MBSs couplings ($\lambda=\lambda'=0$) in Eq.~(\ref{eq:B14}) and Eq.~(\ref{eq:E1}) allows us to recover the well-known Hubbard solution for Green's function of QD~\cite{HubbardI,Bruus}
\begin{equation}
G_{d_{\sigma}d_{\sigma}}^{r}(\omega) = \frac{1-\langle n_{d_{\bar{\sigma}}} \rangle}{\omega^{+}-\varepsilon_{d\sigma}+\imath \Gamma} + \frac{\langle n_{d_{\bar{\sigma}}} \rangle}{\omega^{+}-\varepsilon_{d\sigma}-U+\imath \Gamma}. \label{eq:HbSolution}
\end{equation}


\begin{thebibliography}{54}%
	\makeatletter
	\providecommand \@ifxundefined [1]{%
		\@ifx{#1\undefined}
	}%
	\providecommand \@ifnum [1]{%
		\ifnum #1\expandafter \@firstoftwo
		\else \expandafter \@secondoftwo
		\fi
	}%
	\providecommand \@ifx [1]{%
		\ifx #1\expandafter \@firstoftwo
		\else \expandafter \@secondoftwo
		\fi
	}%
	\providecommand \natexlab [1]{#1}%
	\providecommand \enquote  [1]{``#1''}%
	\providecommand \bibnamefont  [1]{#1}%
	\providecommand \bibfnamefont [1]{#1}%
	\providecommand \citenamefont [1]{#1}%
	\providecommand \href@noop [0]{\@secondoftwo}%
	\providecommand \href [0]{\begingroup \@sanitize@url \@href}%
	\providecommand \@href[1]{\@@startlink{#1}\@@href}%
	\providecommand \@@href[1]{\endgroup#1\@@endlink}%
	\providecommand \@sanitize@url [0]{\catcode `\\12\catcode `\$12\catcode
		`\&12\catcode `\#12\catcode `\^12\catcode `\_12\catcode `\%12\relax}%
	\providecommand \@@startlink[1]{}%
	\providecommand \@@endlink[0]{}%
	\providecommand \url  [0]{\begingroup\@sanitize@url \@url }%
	\providecommand \@url [1]{\endgroup\@href {#1}{\urlprefix }}%
	\providecommand \urlprefix  [0]{URL }%
	\providecommand \Eprint [0]{\href }%
	\providecommand \doibase [0]{http://dx.doi.org/}%
	\providecommand \selectlanguage [0]{\@gobble}%
	\providecommand \bibinfo  [0]{\@secondoftwo}%
	\providecommand \bibfield  [0]{\@secondoftwo}%
	\providecommand \translation [1]{[#1]}%
	\providecommand \BibitemOpen [0]{}%
	\providecommand \bibitemStop [0]{}%
	\providecommand \bibitemNoStop [0]{.\EOS\space}%
	\providecommand \EOS [0]{\spacefactor3000\relax}%
	\providecommand \BibitemShut  [1]{\csname bibitem#1\endcsname}%
	\let\auto@bib@innerbib\@empty
	\bibitem [{\citenamefont {Aguado}(2017)}]{ReviewAguado}%
	\BibitemOpen
	\bibfield  {author} {\bibinfo {author} {\bibfnamefont {R.}~\bibnamefont
			{Aguado}},\ }\href {\doibase 10.1393/ncr/i2017-10141-9} {\bibfield  {journal}
		{\bibinfo  {journal} {Riv. Nuovo Cimento}\ }\textbf {\bibinfo {volume}
			{40}},\ \bibinfo {pages} {523} (\bibinfo {year} {2017})}\BibitemShut
	{NoStop}%
	\bibitem [{\citenamefont {Kitaev}(2001)}]{kitaev}%
	\BibitemOpen
	\bibfield  {author} {\bibinfo {author} {\bibfnamefont {A.~Y.}\ \bibnamefont
			{Kitaev}},\ }\href {http://stacks.iop.org/1063-7869/44/i=10S/a=S29}
	{\bibfield  {journal} {\bibinfo  {journal} {Physics-Uspekhi}\ }\textbf
		{\bibinfo {volume} {44}},\ \bibinfo {pages} {131} (\bibinfo {year}
		{2001})}\BibitemShut {NoStop}%
	\bibitem [{\citenamefont {Nayak}\ \emph {et~al.}(2008)\citenamefont {Nayak},
		\citenamefont {Simon}, \citenamefont {Stern}, \citenamefont {Freedman},\ and\
		\citenamefont {Das~Sarma}}]{RevNonabelian}%
	\BibitemOpen
	\bibfield  {author} {\bibinfo {author} {\bibfnamefont {C.}~\bibnamefont
			{Nayak}}, \bibinfo {author} {\bibfnamefont {S.~H.}\ \bibnamefont {Simon}},
		\bibinfo {author} {\bibfnamefont {A.}~\bibnamefont {Stern}}, \bibinfo
		{author} {\bibfnamefont {M.}~\bibnamefont {Freedman}}, \ and\ \bibinfo
		{author} {\bibfnamefont {S.}~\bibnamefont {Das~Sarma}},\ }\href {\doibase
		10.1103/RevModPhys.80.1083} {\bibfield  {journal} {\bibinfo  {journal} {Rev.
				Mod. Phys.}\ }\textbf {\bibinfo {volume} {80}},\ \bibinfo {pages} {1083}
		(\bibinfo {year} {2008})}\BibitemShut {NoStop}%
	\bibitem [{\citenamefont {Alicea}(2012)}]{Alicea}%
	\BibitemOpen
	\bibfield  {author} {\bibinfo {author} {\bibfnamefont {J.}~\bibnamefont
			{Alicea}},\ }\href {http://stacks.iop.org/0034-4885/75/i=7/a=076501}
	{\bibfield  {journal} {\bibinfo  {journal} {Reports on Progress in Physics}\
		}\textbf {\bibinfo {volume} {75}},\ \bibinfo {pages} {076501} (\bibinfo
		{year} {2012})}\BibitemShut {NoStop}%
	\bibitem [{\citenamefont {Elliott}\ and\ \citenamefont
		{Franz}(2015)}]{RevMajorana}%
	\BibitemOpen
	\bibfield  {author} {\bibinfo {author} {\bibfnamefont {S.~R.}\ \bibnamefont
			{Elliott}}\ and\ \bibinfo {author} {\bibfnamefont {M.}~\bibnamefont
			{Franz}},\ }\href {\doibase 10.1103/RevModPhys.87.137} {\bibfield  {journal}
		{\bibinfo  {journal} {Rev. Mod. Phys.}\ }\textbf {\bibinfo {volume} {87}},\
		\bibinfo {pages} {137} (\bibinfo {year} {2015})}\BibitemShut {NoStop}%
	\bibitem [{\citenamefont {Sarma}\ \emph {et~al.}(2015)\citenamefont {Sarma},
		\citenamefont {Freedman},\ and\ \citenamefont {Nayak}}]{ReviewDasSarma}%
	\BibitemOpen
	\bibfield  {author} {\bibinfo {author} {\bibfnamefont {S.~D.}\ \bibnamefont
			{Sarma}}, \bibinfo {author} {\bibfnamefont {M.}~\bibnamefont {Freedman}}, \
		and\ \bibinfo {author} {\bibfnamefont {C.}~\bibnamefont {Nayak}},\ }\href
	{\doibase 10.1038/npjqi.2015.1} {\bibfield  {journal} {\bibinfo  {journal}
			{Npj Quantum Information}\ }\textbf {\bibinfo {volume} {1}},\ \bibinfo
		{pages} {15001} (\bibinfo {year} {2015})}\BibitemShut {NoStop}%
	\bibitem [{\citenamefont {Goldstein}\ and\ \citenamefont
		{Chamon}(2011)}]{Decaymemories}%
	\BibitemOpen
	\bibfield  {author} {\bibinfo {author} {\bibfnamefont {G.}~\bibnamefont
			{Goldstein}}\ and\ \bibinfo {author} {\bibfnamefont {C.}~\bibnamefont
			{Chamon}},\ }\href {\doibase 10.1103/PhysRevB.84.205109} {\bibfield
		{journal} {\bibinfo  {journal} {Phys. Rev. B}\ }\textbf {\bibinfo {volume}
			{84}},\ \bibinfo {pages} {205109} (\bibinfo {year} {2011})}\BibitemShut
	{NoStop}%
	\bibitem [{\citenamefont {Budich}\ \emph {et~al.}(2012)\citenamefont {Budich},
		\citenamefont {Walter},\ and\ \citenamefont
		{Trauzettel}}]{Failureprotection}%
	\BibitemOpen
	\bibfield  {author} {\bibinfo {author} {\bibfnamefont {J.~C.}\ \bibnamefont
			{Budich}}, \bibinfo {author} {\bibfnamefont {S.}~\bibnamefont {Walter}}, \
		and\ \bibinfo {author} {\bibfnamefont {B.}~\bibnamefont {Trauzettel}},\
	}\href {\doibase 10.1103/PhysRevB.85.121405} {\bibfield  {journal} {\bibinfo
			{journal} {Phys. Rev. B}\ }\textbf {\bibinfo {volume} {85}},\ \bibinfo
		{pages} {121405} (\bibinfo {year} {2012})}\BibitemShut {NoStop}%
	\bibitem [{\citenamefont {Lutchyn}\ \emph {et~al.}(2010)\citenamefont
		{Lutchyn}, \citenamefont {Sau},\ and\ \citenamefont {Das~Sarma}}]{Nanowire1}%
	\BibitemOpen
	\bibfield  {author} {\bibinfo {author} {\bibfnamefont {R.~M.}\ \bibnamefont
			{Lutchyn}}, \bibinfo {author} {\bibfnamefont {J.~D.}\ \bibnamefont {Sau}}, \
		and\ \bibinfo {author} {\bibfnamefont {S.}~\bibnamefont {Das~Sarma}},\ }\href
	{\doibase 10.1103/PhysRevLett.105.077001} {\bibfield  {journal} {\bibinfo
			{journal} {Phys. Rev. Lett.}\ }\textbf {\bibinfo {volume} {105}},\ \bibinfo
		{pages} {077001} (\bibinfo {year} {2010})}\BibitemShut {NoStop}%
	\bibitem [{\citenamefont {Oreg}\ \emph {et~al.}(2010)\citenamefont {Oreg},
		\citenamefont {Refael},\ and\ \citenamefont {von Oppen}}]{Nanowire2}%
	\BibitemOpen
	\bibfield  {author} {\bibinfo {author} {\bibfnamefont {Y.}~\bibnamefont
			{Oreg}}, \bibinfo {author} {\bibfnamefont {G.}~\bibnamefont {Refael}}, \ and\
		\bibinfo {author} {\bibfnamefont {F.}~\bibnamefont {von Oppen}},\ }\href
	{\doibase 10.1103/PhysRevLett.105.177002} {\bibfield  {journal} {\bibinfo
			{journal} {Phys. Rev. Lett.}\ }\textbf {\bibinfo {volume} {105}},\ \bibinfo
		{pages} {177002} (\bibinfo {year} {2010})}\BibitemShut {NoStop}%
	\bibitem [{\citenamefont {Mourik}\ \emph {et~al.}(2012)\citenamefont {Mourik},
		\citenamefont {Zuo}, \citenamefont {Frolov}, \citenamefont {Plissard},
		\citenamefont {Bakkers},\ and\ \citenamefont {Kouwenhoven}}]{mourik2012}%
	\BibitemOpen
	\bibfield  {author} {\bibinfo {author} {\bibfnamefont {V.}~\bibnamefont
			{Mourik}}, \bibinfo {author} {\bibfnamefont {K.}~\bibnamefont {Zuo}},
		\bibinfo {author} {\bibfnamefont {S.~M.}\ \bibnamefont {Frolov}}, \bibinfo
		{author} {\bibfnamefont {S.~R.}\ \bibnamefont {Plissard}}, \bibinfo {author}
		{\bibfnamefont {E.~P. A.~M.}\ \bibnamefont {Bakkers}}, \ and\ \bibinfo
		{author} {\bibfnamefont {L.~P.}\ \bibnamefont {Kouwenhoven}},\ }\href
	{\doibase 10.1126/science.1222360} {\bibfield  {journal} {\bibinfo  {journal}
			{Science}\ }\textbf {\bibinfo {volume} {336}},\ \bibinfo {pages} {1003}
		(\bibinfo {year} {2012})}\BibitemShut {NoStop}%
	\bibitem [{\citenamefont {Albrecht}\ \emph {et~al.}(2016)\citenamefont
		{Albrecht}, \citenamefont {Higginbotham}, \citenamefont {Madsen},
		\citenamefont {Kuemmeth}, \citenamefont {Jespersen}, \citenamefont
		{Nyg{\aa}rd}, \citenamefont {Krogstrup},\ and\ \citenamefont
		{Marcus}}]{albrecht2016exponential}%
	\BibitemOpen
	\bibfield  {author} {\bibinfo {author} {\bibfnamefont {S.~M.}\ \bibnamefont
			{Albrecht}}, \bibinfo {author} {\bibfnamefont {A.}~\bibnamefont
			{Higginbotham}}, \bibinfo {author} {\bibfnamefont {M.}~\bibnamefont
			{Madsen}}, \bibinfo {author} {\bibfnamefont {F.}~\bibnamefont {Kuemmeth}},
		\bibinfo {author} {\bibfnamefont {T.~S.}\ \bibnamefont {Jespersen}}, \bibinfo
		{author} {\bibfnamefont {J.}~\bibnamefont {Nyg{\aa}rd}}, \bibinfo {author}
		{\bibfnamefont {P.}~\bibnamefont {Krogstrup}}, \ and\ \bibinfo {author}
		{\bibfnamefont {C.}~\bibnamefont {Marcus}},\ }\href {\doibase
		10.1038/nature17162} {\bibfield  {journal} {\bibinfo  {journal} {Nature}\
		}\textbf {\bibinfo {volume} {531}},\ \bibinfo {pages} {206} (\bibinfo {year}
		{2016})}\BibitemShut {NoStop}%
	\bibitem [{\citenamefont {Deng}\ \emph {et~al.}(2016)\citenamefont {Deng},
		\citenamefont {Vaitiekenas}, \citenamefont {Hansen}, \citenamefont {Danon},
		\citenamefont {Leijnse}, \citenamefont {Flensberg}, \citenamefont {Nyg{\r
				a}rd}, \citenamefont {Krogstrup},\ and\ \citenamefont {Marcus}}]{Deng2016}%
	\BibitemOpen
	\bibfield  {author} {\bibinfo {author} {\bibfnamefont {M.~T.}\ \bibnamefont
			{Deng}}, \bibinfo {author} {\bibfnamefont {S.}~\bibnamefont {Vaitiekenas}},
		\bibinfo {author} {\bibfnamefont {E.~B.}\ \bibnamefont {Hansen}}, \bibinfo
		{author} {\bibfnamefont {J.}~\bibnamefont {Danon}}, \bibinfo {author}
		{\bibfnamefont {M.}~\bibnamefont {Leijnse}}, \bibinfo {author} {\bibfnamefont
			{K.}~\bibnamefont {Flensberg}}, \bibinfo {author} {\bibfnamefont
			{J.}~\bibnamefont {Nyg{\r a}rd}}, \bibinfo {author} {\bibfnamefont
			{P.}~\bibnamefont {Krogstrup}}, \ and\ \bibinfo {author} {\bibfnamefont
			{C.~M.}\ \bibnamefont {Marcus}},\ }\href {\doibase 10.1126/science.aaf3961}
	{\bibfield  {journal} {\bibinfo  {journal} {Science}\ }\textbf {\bibinfo
			{volume} {354}},\ \bibinfo {pages} {1557} (\bibinfo {year}
		{2016})}\BibitemShut {NoStop}%
	\bibitem [{\citenamefont {Deng}\ \emph {et~al.}(2018)\citenamefont {Deng},
		\citenamefont {Vaitiek\ifmmode~\dot{e}\else \.{e}\fi{}nas}, \citenamefont
		{Prada}, \citenamefont {San-Jose}, \citenamefont {Nyg\aa{}rd}, \citenamefont
		{Krogstrup}, \citenamefont {Aguado},\ and\ \citenamefont
		{Marcus}}]{AguadoExperiment}%
	\BibitemOpen
	\bibfield  {author} {\bibinfo {author} {\bibfnamefont {M.-T.}\ \bibnamefont
			{Deng}}, \bibinfo {author} {\bibfnamefont {S.}~\bibnamefont
			{Vaitiek\ifmmode~\dot{e}\else \.{e}\fi{}nas}}, \bibinfo {author}
		{\bibfnamefont {E.}~\bibnamefont {Prada}}, \bibinfo {author} {\bibfnamefont
			{P.}~\bibnamefont {San-Jose}}, \bibinfo {author} {\bibfnamefont
			{J.}~\bibnamefont {Nyg\aa{}rd}}, \bibinfo {author} {\bibfnamefont
			{P.}~\bibnamefont {Krogstrup}}, \bibinfo {author} {\bibfnamefont
			{R.}~\bibnamefont {Aguado}}, \ and\ \bibinfo {author} {\bibfnamefont {C.~M.}\
			\bibnamefont {Marcus}},\ }\href {\doibase 10.1103/PhysRevB.98.085125}
	{\bibfield  {journal} {\bibinfo  {journal} {Phys. Rev. B}\ }\textbf {\bibinfo
			{volume} {98}},\ \bibinfo {pages} {085125} (\bibinfo {year}
		{2018})}\BibitemShut {NoStop}%
	\bibitem [{\citenamefont {Zhang}\ \emph {et~al.}(2018)\citenamefont {Zhang},
		\citenamefont {Liu}, \citenamefont {Gazibegovic}, \citenamefont {Xu},
		\citenamefont {Logan}, \citenamefont {Wang}, \citenamefont {van Loo},
		\citenamefont {Bommer}, \citenamefont {de~Moor}, \citenamefont {Car},
		\citenamefont {Op~het Veld}, \citenamefont {van Veldhoven}, \citenamefont
		{Koelling}, \citenamefont {Verheijen}, \citenamefont {Pendharkar},
		\citenamefont {Pennachio}, \citenamefont {Shojaei}, \citenamefont {Lee},
		\citenamefont {Palmstr\o{}m}, \citenamefont {Bakkers}, \citenamefont
		{Sarma},\ and\ \citenamefont {Kouwenhoven}}]{Quantized}%
	\BibitemOpen
	\bibfield  {author} {\bibinfo {author} {\bibfnamefont {H.}~\bibnamefont
			{Zhang}}, \bibinfo {author} {\bibfnamefont {C.-X.}\ \bibnamefont {Liu}},
		\bibinfo {author} {\bibfnamefont {S.}~\bibnamefont {Gazibegovic}}, \bibinfo
		{author} {\bibfnamefont {D.}~\bibnamefont {Xu}}, \bibinfo {author}
		{\bibfnamefont {J.~A.}\ \bibnamefont {Logan}}, \bibinfo {author}
		{\bibfnamefont {G.}~\bibnamefont {Wang}}, \bibinfo {author} {\bibfnamefont
			{N.}~\bibnamefont {van Loo}}, \bibinfo {author} {\bibfnamefont {J.~D.~S.}\
			\bibnamefont {Bommer}}, \bibinfo {author} {\bibfnamefont {M.~W.~A.}\
			\bibnamefont {de~Moor}}, \bibinfo {author} {\bibfnamefont {D.}~\bibnamefont
			{Car}}, \bibinfo {author} {\bibfnamefont {R.~L.~M.}\ \bibnamefont {Op~het
				Veld}}, \bibinfo {author} {\bibfnamefont {P.~J.}\ \bibnamefont {van
				Veldhoven}}, \bibinfo {author} {\bibfnamefont {S.}~\bibnamefont {Koelling}},
		\bibinfo {author} {\bibfnamefont {M.~A.}\ \bibnamefont {Verheijen}}, \bibinfo
		{author} {\bibfnamefont {M.}~\bibnamefont {Pendharkar}}, \bibinfo {author}
		{\bibfnamefont {D.~J.}\ \bibnamefont {Pennachio}}, \bibinfo {author}
		{\bibfnamefont {B.}~\bibnamefont {Shojaei}}, \bibinfo {author} {\bibfnamefont
			{J.~S.}\ \bibnamefont {Lee}}, \bibinfo {author} {\bibfnamefont {C.~J.}\
			\bibnamefont {Palmstr\o{}m}}, \bibinfo {author} {\bibfnamefont {E.~P. A.~M.}\
			\bibnamefont {Bakkers}}, \bibinfo {author} {\bibfnamefont {S.~D.}\
			\bibnamefont {Sarma}}, \ and\ \bibinfo {author} {\bibfnamefont {L.~P.}\
			\bibnamefont {Kouwenhoven}},\ }\href {\doibase 10.1038/nature26142}
	{\bibfield  {journal} {\bibinfo  {journal} {Nature}\ }\textbf {\bibinfo
			{volume} {556}},\ \bibinfo {pages} {74} (\bibinfo {year} {2018})}\BibitemShut
	{NoStop}%
	\bibitem [{\citenamefont {Bagrets}\ and\ \citenamefont
		{Altland}(2012)}]{Disorder}%
	\BibitemOpen
	\bibfield  {author} {\bibinfo {author} {\bibfnamefont {D.}~\bibnamefont
			{Bagrets}}\ and\ \bibinfo {author} {\bibfnamefont {A.}~\bibnamefont
			{Altland}},\ }\href {\doibase 10.1103/PhysRevLett.109.227005} {\bibfield
		{journal} {\bibinfo  {journal} {Phys. Rev. Lett.}\ }\textbf {\bibinfo
			{volume} {109}},\ \bibinfo {pages} {227005} (\bibinfo {year}
		{2012})}\BibitemShut {NoStop}%
	\bibitem [{\citenamefont {Cronenwett}\ \emph {et~al.}(1998)\citenamefont
		{Cronenwett}, \citenamefont {Oosterkamp},\ and\ \citenamefont
		{Kouwenhoven}}]{Kondo1}%
	\BibitemOpen
	\bibfield  {author} {\bibinfo {author} {\bibfnamefont {S.~M.}\ \bibnamefont
			{Cronenwett}}, \bibinfo {author} {\bibfnamefont {T.~H.}\ \bibnamefont
			{Oosterkamp}}, \ and\ \bibinfo {author} {\bibfnamefont {L.~P.}\ \bibnamefont
			{Kouwenhoven}},\ }\href {\doibase 10.1126/science.281.5376.540} {\bibfield
		{journal} {\bibinfo  {journal} {Science}\ }\textbf {\bibinfo {volume}
			{281}},\ \bibinfo {pages} {540} (\bibinfo {year} {1998})}\BibitemShut
	{NoStop}%
	\bibitem [{\citenamefont {Goldhaber-Gordon}\ \emph {et~al.}(1998)\citenamefont
		{Goldhaber-Gordon}, \citenamefont {Shtrikman}, \citenamefont {Mahalu},
		\citenamefont {Abusch-Magder}, \citenamefont {Meirav},\ and\ \citenamefont
		{Kastner}}]{Kondo2}%
	\BibitemOpen
	\bibfield  {author} {\bibinfo {author} {\bibfnamefont {D.}~\bibnamefont
			{Goldhaber-Gordon}}, \bibinfo {author} {\bibfnamefont {H.}~\bibnamefont
			{Shtrikman}}, \bibinfo {author} {\bibfnamefont {D.}~\bibnamefont {Mahalu}},
		\bibinfo {author} {\bibfnamefont {D.}~\bibnamefont {Abusch-Magder}}, \bibinfo
		{author} {\bibfnamefont {U.}~\bibnamefont {Meirav}}, \ and\ \bibinfo {author}
		{\bibfnamefont {M.~A.}\ \bibnamefont {Kastner}},\ }\href {\doibase
		10.1038/34373} {\bibfield  {journal} {\bibinfo  {journal} {Nature (London)}\
		}\textbf {\bibinfo {volume} {391}},\ \bibinfo {pages} {156} (\bibinfo {year}
		{1998})}\BibitemShut {NoStop}%
	\bibitem [{\citenamefont {Kells}\ \emph {et~al.}(2012)\citenamefont {Kells},
		\citenamefont {Meidan},\ and\ \citenamefont {Brouwer}}]{ABS1}%
	\BibitemOpen
	\bibfield  {author} {\bibinfo {author} {\bibfnamefont {G.}~\bibnamefont
			{Kells}}, \bibinfo {author} {\bibfnamefont {D.}~\bibnamefont {Meidan}}, \
		and\ \bibinfo {author} {\bibfnamefont {P.~W.}\ \bibnamefont {Brouwer}},\
	}\href {\doibase 10.1103/PhysRevB.86.100503} {\bibfield  {journal} {\bibinfo
			{journal} {Phys. Rev. B}\ }\textbf {\bibinfo {volume} {86}},\ \bibinfo
		{pages} {100503} (\bibinfo {year} {2012})}\BibitemShut {NoStop}%
	\bibitem [{\citenamefont {Lee}\ \emph {et~al.}(2013)\citenamefont {Lee},
		\citenamefont {Jiang}, \citenamefont {Houzet}, \citenamefont {Lieber},\ and\
		\citenamefont {De~Franceschi}}]{ABSLee}%
	\BibitemOpen
	\bibfield  {author} {\bibinfo {author} {\bibfnamefont {E.~J.~H.}\
			\bibnamefont {Lee}}, \bibinfo {author} {\bibfnamefont {X.}~\bibnamefont
			{Jiang}}, \bibinfo {author} {\bibfnamefont {M.}~\bibnamefont {Houzet}},
		\bibinfo {author} {\bibfnamefont {C.~M.}\ \bibnamefont {Lieber}}, \ and\
		\bibinfo {author} {\bibfnamefont {S.}~\bibnamefont {De~Franceschi}},\ }\href
	{\doibase 10.1038/nnano.2013.267} {\bibfield  {journal} {\bibinfo  {journal}
			{Nature Nanotechnology}\ }\textbf {\bibinfo {volume} {9}},\ \bibinfo {pages}
		{79} (\bibinfo {year} {2013})}\BibitemShut {NoStop}%
	\bibitem [{\citenamefont {Liu}\ \emph {et~al.}(2017)\citenamefont {Liu},
		\citenamefont {Sau}, \citenamefont {Stanescu},\ and\ \citenamefont
		{Das~Sarma}}]{ABS2}%
	\BibitemOpen
	\bibfield  {author} {\bibinfo {author} {\bibfnamefont {C.-X.}\ \bibnamefont
			{Liu}}, \bibinfo {author} {\bibfnamefont {J.~D.}\ \bibnamefont {Sau}},
		\bibinfo {author} {\bibfnamefont {T.~D.}\ \bibnamefont {Stanescu}}, \ and\
		\bibinfo {author} {\bibfnamefont {S.}~\bibnamefont {Das~Sarma}},\ }\href
	{\doibase 10.1103/PhysRevB.96.075161} {\bibfield  {journal} {\bibinfo
			{journal} {Phys. Rev. B}\ }\textbf {\bibinfo {volume} {96}},\ \bibinfo
		{pages} {075161} (\bibinfo {year} {2017})}\BibitemShut {NoStop}%
	\bibitem [{\citenamefont {Moore}\ \emph
		{et~al.}(2018{\natexlab{a}})\citenamefont {Moore}, \citenamefont {Stanescu},\
		and\ \citenamefont {Tewari}}]{psABS1}%
	\BibitemOpen
	\bibfield  {author} {\bibinfo {author} {\bibfnamefont {C.}~\bibnamefont
			{Moore}}, \bibinfo {author} {\bibfnamefont {T.~D.}\ \bibnamefont {Stanescu}},
		\ and\ \bibinfo {author} {\bibfnamefont {S.}~\bibnamefont {Tewari}},\ }\href
	{\doibase 10.1103/PhysRevB.97.165302} {\bibfield  {journal} {\bibinfo
			{journal} {Phys. Rev. B}\ }\textbf {\bibinfo {volume} {97}},\ \bibinfo
		{pages} {165302} (\bibinfo {year} {2018}{\natexlab{a}})}\BibitemShut
	{NoStop}%
	\bibitem [{\citenamefont {Moore}\ \emph
		{et~al.}(2018{\natexlab{b}})\citenamefont {Moore}, \citenamefont {Zeng},
		\citenamefont {Stanescu},\ and\ \citenamefont {Tewari}}]{psABS2}%
	\BibitemOpen
	\bibfield  {author} {\bibinfo {author} {\bibfnamefont {C.}~\bibnamefont
			{Moore}}, \bibinfo {author} {\bibfnamefont {C.}~\bibnamefont {Zeng}},
		\bibinfo {author} {\bibfnamefont {T.~D.}\ \bibnamefont {Stanescu}}, \ and\
		\bibinfo {author} {\bibfnamefont {S.}~\bibnamefont {Tewari}},\ }\href@noop {}
	{\bibfield  {journal} {\bibinfo  {journal} {ArXiv e-prints}\ } (\bibinfo
		{year} {2018}{\natexlab{b}})},\ \Eprint {http://arxiv.org/abs/1804.03164}
	{arXiv:1804.03164 [cond-mat.mes-hall]} \BibitemShut {NoStop}%
	\bibitem [{\citenamefont {Cayao}\ \emph {et~al.}(2018)\citenamefont {Cayao},
		\citenamefont {Black-Schaffer}, \citenamefont {Prada},\ and\ \citenamefont
		{Aguado}}]{Cayao2018}%
	\BibitemOpen
	\bibfield  {author} {\bibinfo {author} {\bibfnamefont {J.}~\bibnamefont
			{Cayao}}, \bibinfo {author} {\bibfnamefont {A.~M.}\ \bibnamefont
			{Black-Schaffer}}, \bibinfo {author} {\bibfnamefont {E.}~\bibnamefont
			{Prada}}, \ and\ \bibinfo {author} {\bibfnamefont {R.}~\bibnamefont
			{Aguado}},\ }\href {\doibase 10.3762/bjnano.9.127} {\bibfield  {journal}
		{\bibinfo  {journal} {Beilstein Journal of Nanotechnology}\ }\textbf
		{\bibinfo {volume} {9}},\ \bibinfo {pages} {1339} (\bibinfo {year}
		{2018})}\BibitemShut {NoStop}%
	\bibitem [{\citenamefont {Subhajit}\ and\ \citenamefont {Colin}(2018)}]{YSR}%
	\BibitemOpen
	\bibfield  {author} {\bibinfo {author} {\bibfnamefont {P.}~\bibnamefont
			{Subhajit}}\ and\ \bibinfo {author} {\bibfnamefont {B.}~\bibnamefont
			{Colin}},\ }\href {\doibase 10.1038/s41598-018-30346-4} {\bibfield  {journal}
		{\bibinfo  {journal} {Scientific Reports}\ }\textbf {\bibinfo {volume} {8}},\
		\bibinfo {pages} {11949} (\bibinfo {year} {2018})}\BibitemShut {NoStop}%
	\bibitem [{\citenamefont {Cayao}\ \emph {et~al.}(2015)\citenamefont {Cayao},
		\citenamefont {Prada}, \citenamefont {San-Jose},\ and\ \citenamefont
		{Aguado}}]{SNSjunction}%
	\BibitemOpen
	\bibfield  {author} {\bibinfo {author} {\bibfnamefont {J.}~\bibnamefont
			{Cayao}}, \bibinfo {author} {\bibfnamefont {E.}~\bibnamefont {Prada}},
		\bibinfo {author} {\bibfnamefont {P.}~\bibnamefont {San-Jose}}, \ and\
		\bibinfo {author} {\bibfnamefont {R.}~\bibnamefont {Aguado}},\ }\href
	{\doibase 10.1103/PhysRevB.91.024514} {\bibfield  {journal} {\bibinfo
			{journal} {Phys. Rev. B}\ }\textbf {\bibinfo {volume} {91}},\ \bibinfo
		{pages} {024514} (\bibinfo {year} {2015})}\BibitemShut {NoStop}%
	\bibitem [{\citenamefont {Cayao}\ \emph {et~al.}(2017)\citenamefont {Cayao},
		\citenamefont {San-Jose}, \citenamefont {Black-Schaffer}, \citenamefont
		{Aguado},\ and\ \citenamefont {Prada}}]{Josepheson}%
	\BibitemOpen
	\bibfield  {author} {\bibinfo {author} {\bibfnamefont {J.}~\bibnamefont
			{Cayao}}, \bibinfo {author} {\bibfnamefont {P.}~\bibnamefont {San-Jose}},
		\bibinfo {author} {\bibfnamefont {A.~M.}\ \bibnamefont {Black-Schaffer}},
		\bibinfo {author} {\bibfnamefont {R.}~\bibnamefont {Aguado}}, \ and\ \bibinfo
		{author} {\bibfnamefont {E.}~\bibnamefont {Prada}},\ }\href {\doibase
		10.1103/PhysRevB.96.205425} {\bibfield  {journal} {\bibinfo  {journal} {Phys.
				Rev. B}\ }\textbf {\bibinfo {volume} {96}},\ \bibinfo {pages} {205425}
		(\bibinfo {year} {2017})}\BibitemShut {NoStop}%
	\bibitem [{\citenamefont {San-Jose}\ \emph {et~al.}(2013)\citenamefont
		{San-Jose}, \citenamefont {Cayao}, \citenamefont {Prada},\ and\ \citenamefont
		{Aguado}}]{SanJose2013}%
	\BibitemOpen
	\bibfield  {author} {\bibinfo {author} {\bibfnamefont {P.}~\bibnamefont
			{San-Jose}}, \bibinfo {author} {\bibfnamefont {J.}~\bibnamefont {Cayao}},
		\bibinfo {author} {\bibfnamefont {E.}~\bibnamefont {Prada}}, \ and\ \bibinfo
		{author} {\bibfnamefont {R.}~\bibnamefont {Aguado}},\ }\href {\doibase
		10.1088/1367-2630/15/7/075019} {\bibfield  {journal} {\bibinfo  {journal}
			{New Journal of Physics}\ }\textbf {\bibinfo {volume} {15}},\ \bibinfo
		{pages} {075019} (\bibinfo {year} {2013})}\BibitemShut {NoStop}%
	\bibitem [{\citenamefont {Avila}\ \emph {et~al.}(2018)\citenamefont {Avila},
		\citenamefont {Pe\~naranda}, \citenamefont {Prada}, \citenamefont
		{San-Jose},\ and\ \citenamefont {Aguado}}]{nonHermitan}%
	\BibitemOpen
	\bibfield  {author} {\bibinfo {author} {\bibfnamefont {J.}~\bibnamefont
			{Avila}}, \bibinfo {author} {\bibfnamefont {F.}~\bibnamefont {Pe\~naranda}},
		\bibinfo {author} {\bibfnamefont {E.}~\bibnamefont {Prada}}, \bibinfo
		{author} {\bibfnamefont {P.}~\bibnamefont {San-Jose}}, \ and\ \bibinfo
		{author} {\bibfnamefont {R.}~\bibnamefont {Aguado}},\ }\href@noop {}
	{\bibfield  {journal} {\bibinfo  {journal} {ArXiv e-prints}\ } (\bibinfo
		{year} {2018})},\ \Eprint {http://arxiv.org/abs/1807.04677} {arXiv:1807.04677
		[cond-mat.mes-hall]} \BibitemShut {NoStop}%
	\bibitem [{\citenamefont {Hoffman}\ \emph {et~al.}(2017)\citenamefont
		{Hoffman}, \citenamefont {Chevallier}, \citenamefont {Loss},\ and\
		\citenamefont {Klinovaja}}]{Hoffman}%
	\BibitemOpen
	\bibfield  {author} {\bibinfo {author} {\bibfnamefont {S.}~\bibnamefont
			{Hoffman}}, \bibinfo {author} {\bibfnamefont {D.}~\bibnamefont {Chevallier}},
		\bibinfo {author} {\bibfnamefont {D.}~\bibnamefont {Loss}}, \ and\ \bibinfo
		{author} {\bibfnamefont {J.}~\bibnamefont {Klinovaja}},\ }\href {\doibase
		10.1103/PhysRevB.96.045440} {\bibfield  {journal} {\bibinfo  {journal} {Phys.
				Rev. B}\ }\textbf {\bibinfo {volume} {96}},\ \bibinfo {pages} {045440}
		(\bibinfo {year} {2017})}\BibitemShut {NoStop}%
	\bibitem [{\citenamefont {Chevallier}\ \emph {et~al.}(2018)\citenamefont
		{Chevallier}, \citenamefont {Szumniak}, \citenamefont {Hoffman},
		\citenamefont {Loss},\ and\ \citenamefont {Klinovaja}}]{Denis}%
	\BibitemOpen
	\bibfield  {author} {\bibinfo {author} {\bibfnamefont {D.}~\bibnamefont
			{Chevallier}}, \bibinfo {author} {\bibfnamefont {P.}~\bibnamefont
			{Szumniak}}, \bibinfo {author} {\bibfnamefont {S.}~\bibnamefont {Hoffman}},
		\bibinfo {author} {\bibfnamefont {D.}~\bibnamefont {Loss}}, \ and\ \bibinfo
		{author} {\bibfnamefont {J.}~\bibnamefont {Klinovaja}},\ }\href {\doibase
		10.1103/PhysRevB.97.045404} {\bibfield  {journal} {\bibinfo  {journal} {Phys.
				Rev. B}\ }\textbf {\bibinfo {volume} {97}},\ \bibinfo {pages} {045404}
		(\bibinfo {year} {2018})}\BibitemShut {NoStop}%
	\bibitem [{\citenamefont {Prada}\ \emph {et~al.}(2017)\citenamefont {Prada},
		\citenamefont {Aguado},\ and\ \citenamefont {San-Jose}}]{RAguado}%
	\BibitemOpen
	\bibfield  {author} {\bibinfo {author} {\bibfnamefont {E.}~\bibnamefont
			{Prada}}, \bibinfo {author} {\bibfnamefont {R.}~\bibnamefont {Aguado}}, \
		and\ \bibinfo {author} {\bibfnamefont {P.}~\bibnamefont {San-Jose}},\ }\href
	{\doibase 10.1103/PhysRevB.96.085418} {\bibfield  {journal} {\bibinfo
			{journal} {Phys. Rev. B}\ }\textbf {\bibinfo {volume} {96}},\ \bibinfo
		{pages} {085418} (\bibinfo {year} {2017})}\BibitemShut {NoStop}%
	\bibitem [{\citenamefont {Clarke}(2017)}]{DJClarcke}%
	\BibitemOpen
	\bibfield  {author} {\bibinfo {author} {\bibfnamefont {D.~J.}\ \bibnamefont
			{Clarke}},\ }\href {\doibase 10.1103/PhysRevB.96.201109} {\bibfield
		{journal} {\bibinfo  {journal} {Phys. Rev. B}\ }\textbf {\bibinfo {volume}
			{96}},\ \bibinfo {pages} {201109} (\bibinfo {year} {2017})}\BibitemShut
	{NoStop}%
	\bibitem [{\citenamefont {Liu}\ \emph {et~al.}(2018)\citenamefont {Liu},
		\citenamefont {Sau},\ and\ \citenamefont {Das~Sarma}}]{ABS3}%
	\BibitemOpen
	\bibfield  {author} {\bibinfo {author} {\bibfnamefont {C.-X.}\ \bibnamefont
			{Liu}}, \bibinfo {author} {\bibfnamefont {J.~D.}\ \bibnamefont {Sau}}, \ and\
		\bibinfo {author} {\bibfnamefont {S.}~\bibnamefont {Das~Sarma}},\ }\href
	{\doibase 10.1103/PhysRevB.97.214502} {\bibfield  {journal} {\bibinfo
			{journal} {Phys. Rev. B}\ }\textbf {\bibinfo {volume} {97}},\ \bibinfo
		{pages} {214502} (\bibinfo {year} {2018})}\BibitemShut {NoStop}%
	\bibitem [{\citenamefont {Hell}\ \emph {et~al.}(2018)\citenamefont {Hell},
		\citenamefont {Flensberg},\ and\ \citenamefont {Leijnse}}]{ABS4}%
	\BibitemOpen
	\bibfield  {author} {\bibinfo {author} {\bibfnamefont {M.}~\bibnamefont
			{Hell}}, \bibinfo {author} {\bibfnamefont {K.}~\bibnamefont {Flensberg}}, \
		and\ \bibinfo {author} {\bibfnamefont {M.}~\bibnamefont {Leijnse}},\ }\href
	{\doibase 10.1103/PhysRevB.97.161401} {\bibfield  {journal} {\bibinfo
			{journal} {Phys. Rev. B}\ }\textbf {\bibinfo {volume} {97}},\ \bibinfo
		{pages} {161401} (\bibinfo {year} {2018})}\BibitemShut {NoStop}%
	\bibitem [{\citenamefont {Ricco}\ \emph {et~al.}(2018)\citenamefont {Ricco},
		\citenamefont {Campo}, \citenamefont {Shelykh},\ and\ \citenamefont
		{Seridonio}}]{MBSoscillations}%
	\BibitemOpen
	\bibfield  {author} {\bibinfo {author} {\bibfnamefont {L.~S.}\ \bibnamefont
			{Ricco}}, \bibinfo {author} {\bibfnamefont {V.~L.}\ \bibnamefont {Campo}},
		\bibinfo {author} {\bibfnamefont {I.~A.}\ \bibnamefont {Shelykh}}, \ and\
		\bibinfo {author} {\bibfnamefont {A.~C.}\ \bibnamefont {Seridonio}},\ }\href
	{\doibase 10.1103/PhysRevB.98.075142} {\bibfield  {journal} {\bibinfo
			{journal} {Phys. Rev. B}\ }\textbf {\bibinfo {volume} {98}},\ \bibinfo
		{pages} {075142} (\bibinfo {year} {2018})}\BibitemShut {NoStop}%
	\bibitem [{\citenamefont {Hubbard}(1963)}]{HubbardI}%
	\BibitemOpen
	\bibfield  {author} {\bibinfo {author} {\bibfnamefont {J.}~\bibnamefont
			{Hubbard}},\ }\href {\doibase 10.1098/rspa.1963.0204} {\bibfield  {journal}
		{\bibinfo  {journal} {Proc. R. Soc. A}\ }\textbf {\bibinfo {volume} {276}},\
		\bibinfo {pages} {238} (\bibinfo {year} {1963})}\BibitemShut {NoStop}%
	\bibitem [{\citenamefont {Sticlet}\ \emph {et~al.}(2012)\citenamefont
		{Sticlet}, \citenamefont {Bena},\ and\ \citenamefont {Simon}}]{Bena}%
	\BibitemOpen
	\bibfield  {author} {\bibinfo {author} {\bibfnamefont {D.}~\bibnamefont
			{Sticlet}}, \bibinfo {author} {\bibfnamefont {C.}~\bibnamefont {Bena}}, \
		and\ \bibinfo {author} {\bibfnamefont {P.}~\bibnamefont {Simon}},\ }\href
	{\doibase 10.1103/PhysRevLett.108.096802} {\bibfield  {journal} {\bibinfo
			{journal} {Phys. Rev. Lett.}\ }\textbf {\bibinfo {volume} {108}},\ \bibinfo
		{pages} {096802} (\bibinfo {year} {2012})}\BibitemShut {NoStop}%
	\bibitem [{\citenamefont {Bj\"ornson}\ \emph {et~al.}(2015)\citenamefont
		{Bj\"ornson}, \citenamefont {Pershoguba}, \citenamefont {Balatsky},\ and\
		\citenamefont {Black-Schaffer}}]{Bjornson}%
	\BibitemOpen
	\bibfield  {author} {\bibinfo {author} {\bibfnamefont {K.}~\bibnamefont
			{Bj\"ornson}}, \bibinfo {author} {\bibfnamefont {S.~S.}\ \bibnamefont
			{Pershoguba}}, \bibinfo {author} {\bibfnamefont {A.~V.}\ \bibnamefont
			{Balatsky}}, \ and\ \bibinfo {author} {\bibfnamefont {A.~M.}\ \bibnamefont
			{Black-Schaffer}},\ }\href {\doibase 10.1103/PhysRevB.92.214501} {\bibfield
		{journal} {\bibinfo  {journal} {Phys. Rev. B}\ }\textbf {\bibinfo {volume}
			{92}},\ \bibinfo {pages} {214501} (\bibinfo {year} {2015})}\BibitemShut
	{NoStop}%
	\bibitem [{\citenamefont {Ma\'ska}\ and\ \citenamefont
		{Doma\'nski}(2017)}]{Tadeusz}%
	\BibitemOpen
	\bibfield  {author} {\bibinfo {author} {\bibfnamefont {M.~M.}\ \bibnamefont
			{Ma\'ska}}\ and\ \bibinfo {author} {\bibfnamefont {T.}~\bibnamefont
			{Doma\'nski}},\ }\href {\doibase 10.1038/s41598-017-16323-3} {\bibfield
		{journal} {\bibinfo  {journal} {Scientific Reports}\ }\textbf {\bibinfo
			{volume} {7}},\ \bibinfo {pages} {16193} (\bibinfo {year}
		{2017})}\BibitemShut {NoStop}%
	\bibitem [{\citenamefont {Jeon}\ \emph {et~al.}(2017)\citenamefont {Jeon},
		\citenamefont {Xie}, \citenamefont {Li}, \citenamefont {Wang}, \citenamefont
		{Bernevig},\ and\ \citenamefont {Yazdani}}]{Jeon}%
	\BibitemOpen
	\bibfield  {author} {\bibinfo {author} {\bibfnamefont {S.}~\bibnamefont
			{Jeon}}, \bibinfo {author} {\bibfnamefont {Y.}~\bibnamefont {Xie}}, \bibinfo
		{author} {\bibfnamefont {J.}~\bibnamefont {Li}}, \bibinfo {author}
		{\bibfnamefont {Z.}~\bibnamefont {Wang}}, \bibinfo {author} {\bibfnamefont
			{B.~A.}\ \bibnamefont {Bernevig}}, \ and\ \bibinfo {author} {\bibfnamefont
			{A.}~\bibnamefont {Yazdani}},\ }\href {\doibase 10.1126/science.aan3670}
	{\bibfield  {journal} {\bibinfo  {journal} {Science}\ }\textbf {\bibinfo
			{volume} {358}},\ \bibinfo {pages} {772} (\bibinfo {year}
		{2017})}\BibitemShut {NoStop}%
	\bibitem [{\citenamefont {Li}\ \emph {et~al.}(2018)\citenamefont {Li},
		\citenamefont {Jeon}, \citenamefont {Xie}, \citenamefont {Yazdani},\ and\
		\citenamefont {Bernevig}}]{Andrei}%
	\BibitemOpen
	\bibfield  {author} {\bibinfo {author} {\bibfnamefont {J.}~\bibnamefont
			{Li}}, \bibinfo {author} {\bibfnamefont {S.}~\bibnamefont {Jeon}}, \bibinfo
		{author} {\bibfnamefont {Y.}~\bibnamefont {Xie}}, \bibinfo {author}
		{\bibfnamefont {A.}~\bibnamefont {Yazdani}}, \ and\ \bibinfo {author}
		{\bibfnamefont {B.~A.}\ \bibnamefont {Bernevig}},\ }\href {\doibase
		10.1103/PhysRevB.97.125119} {\bibfield  {journal} {\bibinfo  {journal} {Phys.
				Rev. B}\ }\textbf {\bibinfo {volume} {97}},\ \bibinfo {pages} {125119}
		(\bibinfo {year} {2018})}\BibitemShut {NoStop}%
	\bibitem [{\citenamefont {Serina}\ \emph {et~al.}(2018)\citenamefont {Serina},
		\citenamefont {Loss},\ and\ \citenamefont {Klinovaja}}]{Jelenaspinpolarized}%
	\BibitemOpen
	\bibfield  {author} {\bibinfo {author} {\bibfnamefont {M.}~\bibnamefont
			{Serina}}, \bibinfo {author} {\bibfnamefont {D.}~\bibnamefont {Loss}}, \ and\
		\bibinfo {author} {\bibfnamefont {J.}~\bibnamefont {Klinovaja}},\ }\href
	{\doibase 10.1103/PhysRevB.98.035419} {\bibfield  {journal} {\bibinfo
			{journal} {Phys. Rev. B}\ }\textbf {\bibinfo {volume} {98}},\ \bibinfo
		{pages} {035419} (\bibinfo {year} {2018})}\BibitemShut {NoStop}%
	\bibitem [{\citenamefont {Danon}\ \emph {et~al.}(2017)\citenamefont {Danon},
		\citenamefont {Hansen},\ and\ \citenamefont {Flensberg}}]{Danon}%
	\BibitemOpen
	\bibfield  {author} {\bibinfo {author} {\bibfnamefont {J.}~\bibnamefont
			{Danon}}, \bibinfo {author} {\bibfnamefont {E.~B.}\ \bibnamefont {Hansen}}, \
		and\ \bibinfo {author} {\bibfnamefont {K.}~\bibnamefont {Flensberg}},\ }\href
	{\doibase 10.1103/PhysRevB.96.125420} {\bibfield  {journal} {\bibinfo
			{journal} {Phys. Rev. B}\ }\textbf {\bibinfo {volume} {96}},\ \bibinfo
		{pages} {125420} (\bibinfo {year} {2017})}\BibitemShut {NoStop}%
	\bibitem [{\citenamefont {Fleckenstein}\ \emph {et~al.}(2018)\citenamefont
		{Fleckenstein}, \citenamefont {Dom\'{\i}nguez}, \citenamefont
		{Traverso~Ziani},\ and\ \citenamefont {Trauzettel}}]{Fleckenstein}%
	\BibitemOpen
	\bibfield  {author} {\bibinfo {author} {\bibfnamefont {C.}~\bibnamefont
			{Fleckenstein}}, \bibinfo {author} {\bibfnamefont {F.}~\bibnamefont
			{Dom\'{\i}nguez}}, \bibinfo {author} {\bibfnamefont {N.}~\bibnamefont
			{Traverso~Ziani}}, \ and\ \bibinfo {author} {\bibfnamefont {B.}~\bibnamefont
			{Trauzettel}},\ }\href {\doibase 10.1103/PhysRevB.97.155425} {\bibfield
		{journal} {\bibinfo  {journal} {Phys. Rev. B}\ }\textbf {\bibinfo {volume}
			{97}},\ \bibinfo {pages} {155425} (\bibinfo {year} {2018})}\BibitemShut
	{NoStop}%
	\bibitem [{\citenamefont {Anderson}(1961)}]{Anderson}%
	\BibitemOpen
	\bibfield  {author} {\bibinfo {author} {\bibfnamefont {P.~W.}\ \bibnamefont
			{Anderson}},\ }\href {\doibase 10.1103/PhysRev.124.41} {\bibfield  {journal}
		{\bibinfo  {journal} {Phys. Rev.}\ }\textbf {\bibinfo {volume} {124}},\
		\bibinfo {pages} {41} (\bibinfo {year} {1961})}\BibitemShut {NoStop}%
	\bibitem [{\citenamefont {Haug}\ and\ \citenamefont {Jauho}(2008)}]{Jauho}%
	\BibitemOpen
	\bibfield  {author} {\bibinfo {author} {\bibfnamefont {H.}~\bibnamefont
			{Haug}}\ and\ \bibinfo {author} {\bibfnamefont {A.}~\bibnamefont {Jauho}},\
	}\href@noop {} {\emph {\bibinfo {title} {Quantum Kinetics in Transport and
				Optics of Semiconductors}}},\ Springer Series in Solid-State Sciences\
	(\bibinfo  {publisher} {Springer Berlin Heidelberg},\ \bibinfo {year}
	{2008})\BibitemShut {NoStop}%
	\bibitem [{\citenamefont {Bruus}\ and\ \citenamefont
		{Flensberg}(2004)}]{Bruus}%
	\BibitemOpen
	\bibfield  {author} {\bibinfo {author} {\bibfnamefont {H.}~\bibnamefont
			{Bruus}}\ and\ \bibinfo {author} {\bibfnamefont {K.}~\bibnamefont
			{Flensberg}},\ }\href@noop {} {\emph {\bibinfo {title} {Many-Body Quantum
				Theory in Condensed Matter Physics: An Introduction}}},\ Oxford Graduate
	Texts\ (\bibinfo  {publisher} {Oxford University Press},\ \bibinfo {year}
	{2004})\BibitemShut {NoStop}%
	\bibitem [{\citenamefont {Vernek}\ \emph {et~al.}(2014)\citenamefont {Vernek},
		\citenamefont {Penteado}, \citenamefont {Seridonio},\ and\ \citenamefont
		{Egues}}]{Vernek}%
	\BibitemOpen
	\bibfield  {author} {\bibinfo {author} {\bibfnamefont {E.}~\bibnamefont
			{Vernek}}, \bibinfo {author} {\bibfnamefont {P.~H.}\ \bibnamefont
			{Penteado}}, \bibinfo {author} {\bibfnamefont {A.~C.}\ \bibnamefont
			{Seridonio}}, \ and\ \bibinfo {author} {\bibfnamefont {J.~C.}\ \bibnamefont
			{Egues}},\ }\href {\doibase 10.1103/PhysRevB.89.165314} {\bibfield  {journal}
		{\bibinfo  {journal} {Phys. Rev. B}\ }\textbf {\bibinfo {volume} {89}},\
		\bibinfo {pages} {165314} (\bibinfo {year} {2014})}\BibitemShut {NoStop}%
	\bibitem [{\citenamefont {Bara{\'{n}}ski}\ and\ \citenamefont
		{Doma{\'{n}}ski}(2013)}]{Baraski2013}%
	\BibitemOpen
	\bibfield  {author} {\bibinfo {author} {\bibfnamefont {J.}~\bibnamefont
			{Bara{\'{n}}ski}}\ and\ \bibinfo {author} {\bibfnamefont {T.}~\bibnamefont
			{Doma{\'{n}}ski}},\ }\href {\doibase 10.1088/0953-8984/25/43/435305}
	{\bibfield  {journal} {\bibinfo  {journal} {J. Phys.: Cond. Matter}\ }\textbf
		{\bibinfo {volume} {25}},\ \bibinfo {pages} {435305} (\bibinfo {year}
		{2013})}\BibitemShut {NoStop}%
	\bibitem [{\citenamefont {Balatsky}\ \emph {et~al.}(2006)\citenamefont
		{Balatsky}, \citenamefont {Vekhter},\ and\ \citenamefont
		{Zhu}}]{RevModPhysimpuritiesSC}%
	\BibitemOpen
	\bibfield  {author} {\bibinfo {author} {\bibfnamefont {A.~V.}\ \bibnamefont
			{Balatsky}}, \bibinfo {author} {\bibfnamefont {I.}~\bibnamefont {Vekhter}}, \
		and\ \bibinfo {author} {\bibfnamefont {J.-X.}\ \bibnamefont {Zhu}},\ }\href
	{\doibase 10.1103/RevModPhys.78.373} {\bibfield  {journal} {\bibinfo
			{journal} {Rev. Mod. Phys.}\ }\textbf {\bibinfo {volume} {78}},\ \bibinfo
		{pages} {373} (\bibinfo {year} {2006})}\BibitemShut {NoStop}%
	\bibitem [{\citenamefont {Miroshnichenko}\ \emph {et~al.}(2010)\citenamefont
		{Miroshnichenko}, \citenamefont {Flach},\ and\ \citenamefont
		{Kivshar}}]{RevFano}%
	\BibitemOpen
	\bibfield  {author} {\bibinfo {author} {\bibfnamefont {A.~E.}\ \bibnamefont
			{Miroshnichenko}}, \bibinfo {author} {\bibfnamefont {S.}~\bibnamefont
			{Flach}}, \ and\ \bibinfo {author} {\bibfnamefont {Y.~S.}\ \bibnamefont
			{Kivshar}},\ }\href {\doibase 10.1103/RevModPhys.82.2257} {\bibfield
		{journal} {\bibinfo  {journal} {Rev. Mod. Phys.}\ }\textbf {\bibinfo {volume}
			{82}},\ \bibinfo {pages} {2257} (\bibinfo {year} {2010})}\BibitemShut
	{NoStop}%
	\bibitem [{\citenamefont {F\"ul\"op}\ \emph {et~al.}(2015)\citenamefont
		{F\"ul\"op}, \citenamefont {Dom\'{\i}nguez}, \citenamefont {d'Hollosy},
		\citenamefont {Baumgartner}, \citenamefont {Makk}, \citenamefont {Madsen},
		\citenamefont {Guzenko}, \citenamefont {Nyg\aa{}rd}, \citenamefont
		{Sch\"onenberger}, \citenamefont {Levy~Yeyati},\ and\ \citenamefont
		{Csonka}}]{CPS}%
	\BibitemOpen
	\bibfield  {author} {\bibinfo {author} {\bibfnamefont {G.}~\bibnamefont
			{F\"ul\"op}}, \bibinfo {author} {\bibfnamefont {F.}~\bibnamefont
			{Dom\'{\i}nguez}}, \bibinfo {author} {\bibfnamefont {S.}~\bibnamefont
			{d'Hollosy}}, \bibinfo {author} {\bibfnamefont {A.}~\bibnamefont
			{Baumgartner}}, \bibinfo {author} {\bibfnamefont {P.}~\bibnamefont {Makk}},
		\bibinfo {author} {\bibfnamefont {M.~H.}\ \bibnamefont {Madsen}}, \bibinfo
		{author} {\bibfnamefont {V.~A.}\ \bibnamefont {Guzenko}}, \bibinfo {author}
		{\bibfnamefont {J.}~\bibnamefont {Nyg\aa{}rd}}, \bibinfo {author}
		{\bibfnamefont {C.}~\bibnamefont {Sch\"onenberger}}, \bibinfo {author}
		{\bibfnamefont {A.}~\bibnamefont {Levy~Yeyati}}, \ and\ \bibinfo {author}
		{\bibfnamefont {S.}~\bibnamefont {Csonka}},\ }\href {\doibase
		10.1103/PhysRevLett.115.227003} {\bibfield  {journal} {\bibinfo  {journal}
			{Phys. Rev. Lett.}\ }\textbf {\bibinfo {volume} {115}},\ \bibinfo {pages}
		{227003} (\bibinfo {year} {2015})}\BibitemShut {NoStop}%
	\bibitem [{\citenamefont {Schuray}\ \emph {et~al.}(2017)\citenamefont
		{Schuray}, \citenamefont {Weithofer},\ and\ \citenamefont {Recher}}]{Recher}%
	\BibitemOpen
	\bibfield  {author} {\bibinfo {author} {\bibfnamefont {A.}~\bibnamefont
			{Schuray}}, \bibinfo {author} {\bibfnamefont {L.}~\bibnamefont {Weithofer}},
		\ and\ \bibinfo {author} {\bibfnamefont {P.}~\bibnamefont {Recher}},\ }\href
	{\doibase 10.1103/PhysRevB.96.085417} {\bibfield  {journal} {\bibinfo
			{journal} {Phys. Rev. B}\ }\textbf {\bibinfo {volume} {96}},\ \bibinfo
		{pages} {085417} (\bibinfo {year} {2017})}\BibitemShut {NoStop}%
\end{thebibliography}
\end{document}